\documentclass[aps,prd,amsfonts,nofootinbib,supercriptaddress,11pt]{revtex4}
\usepackage{amssymb,amsmath}
\usepackage{epsfig}
\usepackage{comment}
\usepackage{setspace}
\usepackage{natbib}

\newcounter{fig}

\newcommand{\beq}{\begin{equation}}
\newcommand{\eeq}{\end{equation}}
\newcommand{\bea}{\begin{eqnarray}}
\newcommand{\eea}{\end{eqnarray}}

\begin{document}


\vspace{0.5cm}

\title{A novel black hole mimicker: a boson star and a global monopole nonminimally coupled to gravity}

\author{Anja Marunovi\'c$^a$\footnote{anja.marunovic@fer.hr} and Miljenko Murkovi\'c\footnote{miljenkomelo@gmail.com}}
\affiliation{$^a$University of Zagreb,  Faculty of Electrical
Engineering and Computing, Physics Department, Unska 3, 10
000 Zagreb, Croatia }

\begin{abstract} \noindent
A field-theoretic model for a highly compact object that mimicks a black hole is found for the gravitationally interacting system of a boson star and a global monopole which are nonminimally coupled to gravity. According to the strength of the nonlinear gravitational effects and the gravitational backreaction, three distinct coupling regimes are featured: weak, mild and strong. In the strong coupling regime we show that a repulsive monopole stabilizes an attractive boson star and the resulting configuration exhibits large energy density, large (and negative) principal pressures, large compactness, large effective potential, large local forces, and yet exhibits no event horizon. As such a composite system of a boson star and a global monopole represents a convincing microscopic candidate for a \emph{black hole mimicker}.
\end{abstract}


\maketitle

%
\section{Introduction}

The notorious information loss paradox, apart from introducing many heated discussions on the quantum front of physics, has also put in question a singular collapse by which black holes form. Hence the existence of black holes naturally emerges as conceptually undesirable. Mathematically, they certainly do exist as stationary vacuum solutions of the Einstein equations. Physically, however, due to the measurement constraints, it is not yet possible to tell if a final state of a massive star collapse is a black hole or some other very dense compact object. Even though, the theory generally allows for a  nonsingular collapse (by, for example, choosing some proper (usually weird) equation of state), there is only a limited set of physically acceptable black hole alternatives~\cite{Visser1}: quark stars, Q-balls, strange stars, boson stars, gravastars, fuzz-balls and dark stars/quasi-black holes (see references in~\cite{Visser1}). Furthermore, wormholes as black hole mimickers were investigated in~\cite{Zaslavskii}.

The oldest, and accordingly the most studied, astrophysical example based on the \emph{Lagrangian formalism}, is the boson star, which is a compact object built from a self-interacting, gravitationally bound scalar field (on boson stars see, \emph{e.g.},~\cite{Kaup,Ruffini,Colpi,Friedberg,LeePang,Jetzer,Seidel,Liebling,SchunckRev,JetzerRev,LiddleRev}). It is known that boson stars coupled to Einstein's general relativity possess some features that characterize also gravastars (on gravastars see, \emph{e.g.},~\cite{MM,Visser,Bilic,Cattoen,DeBenedictis,Rezzolla,Horvat:Charged,Harko,Horvat:Radial,Visser2}), such as the anisotropy in principal pressures and relatively large compactness ($\mu_{\rm max}= 0.32$). However, no matter how large the self-coupling is, the ordinary boson star cannot attain arbitrarily large compression and as such does not represent a very good black hole mimicker. Furthermore, the principal pressures do not have a de Sitter-like interior  - that is, their principal pressures are always positive at the origin. In our recent work~\cite{moiDES} we extended the analysis of boson stars  and modified the Einstein-Hilbert action by introducing a nonminimal coupling of the scalar field to gravity via the Ricci curvature scalar. 
Even though boson stars with with nonminimal coupling to gravity were already investigated in~\cite{Bij Gleiser}, and also within conformal gravity and its scalar-tensor extensions in~\cite{Conformal}, in~\cite{moiDES} we rather set focus on configurations that resemble more the dark energy stars then the ordinary boson stars, and show that their compactness is significantly larger than that in ordinary boson stars (if matter is not constrained by energy conditions).

A field-theoretic model that we investigate in this paper involves a global monopole and a combined system of a boson star and a global monopole. Global monopoles are extensively studied configurations in the context of cosmology. Namely, they belong to the class of topological defects, whose networks were studied in the 1980s and 1990s as a possible origin of the Universe large scale structure. Modern cosmological observations have ruled out topological defects as the principal seeds for structure formation, albeit a small fraction of cosmic microwave background thermal fluctuations might still originate from topological defects~\cite{Ade:2013xla}.  The simplest field-theoretic realization of the global monopole includes a scalar field theory with an (global) $\mathcal O(3)$ - symmetry which is spontaneously broken to $\mathcal O(2)$ by the vacuum~\cite{Vilenkin}. Within the framework of classical general relativity, the most prominent feature of the global monopole is the gravitationally repulsive core mass~\cite{Lousto}. The idea of the so called topological inflation was also considered in the context of global monopoles due to the existence of de Sitter cores~\cite{Perivoj}. When gravity is modified by introducing a nonminimal coupling however, the locally attractive regions of effective force emerge, thus enabling the existence of bound orbits~\cite{NucamendiBO,NucamendiDM}. Due to all these peculiar features of global monopoles, it seems reasonable to investigate the system consisting of a boson star and a global monopole, the so-called \emph{D-star}.
 Indeed, in this paper we show that a repulsive monopole stabilizes an attractive boson star and the resulting configuration exhibits large energy density, large (and negative) principal pressures, large compactness, large effective potential, large local forces, and yet exhibits no event horizon. As such a composite system of a boson star and a global monopole represents a convincing microscopic candidate for a \emph{black hole mimicker}.

D-stars or topological defect stars are "compact objects with a solid angular deficit, which generalize Q-stars by including a complex scalar field (or a fermion field), the Goldstone field and classical Einstein gravity"~\cite{KineziGravLens}. In Ref.~\cite{KineziFermionD} a fermion D-star is investigated while in Ref.~\cite{KineziBosonD} an analysis of a boson D-star is performed. While fermion D-stars showed yet unresolved issues on the stability, boson D-stars have revealed an attractive features in the context of compact objects. Even though the authors presented an approximate solutions to the gravitational field outside D-stars, their analysis motivates the existence of black holes with deficit solid angle. Furthermore, in Ref.~\cite{KineziMotionD} the motion around D-stars is  investigated while in Ref.~\cite{KineziGravLens}  D-stars as gravitational lenses were considered.

 This paper is organized as follows: in section~\ref{Model} we present a derivation of the Einstein equations for a composite system of a boson star and a global monopole in which both fields are nonminimally coupled to gravity and interact only through the gravitational field. In section~\ref{NGM} we analyze a
 global monopole which is nonminimally coupled to gravity. In subsection~\ref{NGM R1} the results for the global monopole which is minimally coupled to gravity is briefly demonstrated, while the main results for the nonminimally coupled global monopole are  presented in subsection~\ref{NGM R2} including metric functions, energy densities, pressures, compactnesses, effective forces, Newtonian forces and Newtonian forces produces by core masses. In section~\ref{BSGM} we analyze a composite system of a boson star and a global monopole. In~\ref{Weak} we show results for weakly coupled boson star and a global monopole, for which both fields are only slightly affected in the combined system. In~\ref{Mild} results for mild coupling are presented while in~\ref{Strong} we show solutions for the composite system produced as a result of the strong (to extremal) coupling  among a boson star and a global monopole field. In this regime we find a set of parameters for which a good black hole mimicker with the deficit solid angle may form.


\section{The model}\label{Model}
For gravity we take the standard Einstein--Hilbert action
\begin{equation}
S_{\mathrm{EH}}=\int\mathrm{d}x^{4}\sqrt{-g}\frac{R}{16\pi G_{\mathrm{N}}}\,,
\end{equation}
where $G_{\mathrm{N}}$ is the Newton constant, $R$ is the Ricci scalar
and $g$ is the determinant of the metric tensor $g_{\mu\nu}$ which for spherically symmetric systems can be written as
\begin{equation}
g_{\mu\nu}=\mathrm{diag}\left(-e^{\nu(r)},e^{\lambda(r)},r^{2},r^{2}\sin^{2}(\theta)\right)\label{sferna metrika}.
\end{equation}
The action that describes the boson star in the nonminimal setting is:
\begin{equation}
S_{\mathrm{BS}}=\int\mathrm{d}x^{4}\sqrt{-g}\left[-g^{\mu\nu}\left(\nabla_{\mu}\sigma^{*}\right)\left(\nabla_{\nu}\sigma\right)-m^{2}
\left(\sigma^{*}\sigma\right)-\frac{\lambda_{\mathrm{BS}}}{2}\left(\sigma^{*}\sigma\right)^2+
\xi_{\mathrm{BS}}R\left(\sigma^{*}\sigma\right)\right],
\label{action BS}
\end{equation}
where $m$ is the scalar field mass, $\lambda_{\rm BS}$ is the quartic self-interaction and $\sigma^*$ is the complex conjugate of the scalar field $\sigma$.  $\xi_{\mathrm{BS}}$ is the quantity that  measures strength of the coupling between the scalar field $\sigma$ and gravity via Ricci scalar $R$.\\
The action for the nonminimally coupled global monopole is
\begin{equation}
S_{\mathrm{GM}}=\int\mathrm{d}x^{4}\sqrt{-g}\left[-\frac{1}{2}g^{\mu\nu}\left(\nabla_{\mu}\phi^{a}\right)
\left(\nabla_{\nu}\phi^{a}\right)-V\left(\phi^{a}\right)+\frac{\xi_{\mathrm{GM}}}{2}R\left(\phi^{a}\phi^{a}\right)\right],
\label{action GM}
\end{equation}
where $\phi^{a}$, $a=1,2,3$, is a scalar field triplet  with global $\mathcal{O}(3)$ symmetry which is spontaneously
broken to $\mathcal O(2)$ with the simplest symmetry breaking potential
\begin{equation}
V\left(\phi^{a}\right)=\frac{\mu^{2}}{2}\phi^{a}\phi^{a}+\frac{\lambda_{\mathrm{GM}}}{4}
\left(\phi^{a}\phi^{a}\right)^{2}+\frac{\mu^{4}}{4\lambda_{\mathrm{GM}}}\label{potentialA}.
\end{equation}
$\mu$ is the monopole mass term and $\lambda_{\mathrm{GM}}$ is its self-coupling.
The quantity $\xi_{\mathrm{GM}}$ measures the strength
of the coupling between the monopole scalar field and gravity via Ricci scalar. \\
For the boson star field we choose a harmonic time-dependence \emph{Ansatz}
\begin{equation}
\sigma(r,t)=\sigma(r)e^{-i\omega t}\,,\label{ansatz BS}
\end{equation}
while for the global monopole configuration we use the (standard) so-called hedgehog
\emph{Ansatz}:
\begin{equation}
\phi^{a}=\phi(r)\hat{r}^{a}\,,\qquad\hat{r}=\left\{ \sin\theta\cos\varphi,\:\sin\theta\sin\varphi,\:\cos\theta\right\} .\label{ansatz GM}
\end{equation}
With the given \emph{Ansatz} \eqref{ansatz GM} the potential \eqref{potentialA}
can be written as
\begin{equation}
V\left(\phi\right)=\frac{\lambda_{\mathrm{GM}}}{4}\left(\phi^{2}-\phi_{0}^{2}\right)^{2},\label{potentialB}
\end{equation}
where $\phi_{0}^{2}=-\mu^{2}/\lambda_{\mathrm{GM}}$ and characterizes
the energy of the symmetry breaking scale.\\
The energy-momentum tensor of the boson star is obtained by varying its action~(\ref{action BS}) with respect to the  metric tensor $g^{\mu\nu}$ yielding
\begin{alignat}{1}
T_{\mu\nu}^{\mathrm{BS}}=  -\frac{2}{\sqrt{-g}}\frac{\delta S_{\rm BS}}{\delta g^{\mu\nu}} = & 2\delta_{(\mu}^{\alpha}\delta_{\nu)}^{\beta}\left(\nabla_{\alpha}\sigma^{*}\right)\left(\nabla_{\beta}\sigma\right)-g_{\mu\nu}\left[g^{\alpha\beta}\left(\nabla_{\alpha}\sigma^{*}\right)\left(\nabla_{\beta}\sigma\right)+m^{2}\left(\sigma^{*}\sigma\right)+\frac{\lambda_{\mathrm{BS}}}{2}\left(\sigma^{*}\sigma\right)^{2}\right]\nonumber \\
 & -2\xi_{\mathrm{BS}}\left(G_{\mu\nu}+g_{\mu\nu}\square-\nabla_{\mu}\nabla_{\nu}\right)
 \left(\sigma^{*}\sigma\right).\label{energy-momentum tensorBS}
\end{alignat}
Similarly, the energy-momentum tensor for the global monopole field is
\begin{alignat}{1}
T_{\mu\nu}^{\mathrm{GM}}=   -\frac{2}{\sqrt{-g}}\frac{\delta S_{\rm GM}}{\delta g^{\mu\nu}}= & \left(\nabla_{\mu}\phi^{a}\right)\left(\nabla_{\nu}\phi^{a}\right)-g_{\mu\nu}\left[\frac{1}{2}g^{\alpha\beta}\left(\nabla_{\alpha}\phi^{a}\right)\left(\nabla_{\beta}\phi^{a}\right)+V\left(\phi^{a}\right)\right]\nonumber \\
 & -\xi_{\mathrm{GM}}\left(G_{\mu\nu}+g_{\mu\nu}\square-\nabla_{\mu}\nabla_{\nu}\right)\left(\phi^{a}\phi^{a}\right)\,,\label{energy-momentum tensorGM}
\end{alignat}
where $\square=g^{\alpha\beta}\nabla_{\alpha}\nabla_{\beta}$ is the d'Alembertian operator.\\

For the purpose of numerical studies it is
convenient to work with the dimensionless variables/functions. Hence we
make the rescaling upon which all the variables/functions are given in the reduced Planck units
\[
r=\sqrt{8\pi G_{\mathrm{N}}}x\,,\qquad 8\pi G_{\mathrm{N}}\sigma^{2}=\tilde{\sigma}^{2}\,,\qquad8\pi G_{\mathrm{N}}\left(\frac{\phi}{\phi_0}\right)^{2}=\tilde{\phi}^{2}\,,
\]
\[
8\pi G_{\mathrm{N}}R=\tilde{R}\,,\qquad8\pi G_{\mathrm{N}}m^{2}=\tilde{m}^{2}\,,\qquad8\pi G_{\mathrm{N}}\omega^{2}=\tilde{\omega}^{2}\,.
\label{rescalings}
\]
If we identify the components of the energy-momentum tensor as
\beq
T_{\mu\nu}=\mbox{diag}(-\rho,p_r,p_t,p_t)\,,
\eeq
it follows that the dimensionless energy density and pressures are
\beq
\tilde\rho=(8\pi G_{\mathrm{N}})^2\rho,\qquad \tilde{p}_{r,t}=(8\pi G_{\mathrm{N}})^2p_{r,t}.
\eeq
The equations of motions for the boson star and the global monopole fields
are obtained by varying the total action $S_{\mathrm{tot}}=S_{\rm EH}+S_{\rm BS}+S_{\rm GM}$ with respect
to $\sigma^*$ and $\phi^*$, respectively. Using the above rescaling
and \emph{Ansatze} \eqref{ansatz BS} and \eqref{ansatz GM}  we arrive at
\beq
\tilde{\sigma}^{\prime\prime}=-\left(\frac 2x +\frac{\nu^\prime-\lambda^\prime}{2}\right)\tilde{\sigma}^\prime+e^\lambda(\tilde m^2+\lambda_{\mathrm{BS}}\tilde{\sigma}^2-\tilde\omega^2 e^{-\nu}-\xi_{\mathrm{BS}}\tilde R)\tilde{\sigma},
\label{eom BS BSGM}
\eeq
and
\beq
\tilde\phi^{\prime\prime}=-\left(\frac 2x +\frac{\nu^\prime-\lambda^\prime}{2}\right)\tilde\phi^\prime
+e^\lambda\left(\lambda_{\mathrm{GM}}\Delta(\tilde\phi^2 -1)+\frac{2}{x^2}-\xi_{\mathrm{GM}}\tilde R\right)\tilde\phi,
\label{eom GM BSGM}
\eeq
where
\beq
\Delta=8\pi G_N\phi_0^2\,,
\eeq
is the energy of the symmetry breaking scale (in the reduced Planck units) and the meaning is a deficit solid angle.\\

From the first two Einstein equations $G_{\mu\nu}=8\pi G_{\mathrm{N}} T_{\mu\nu}$ we obtain the differential equations for the metric functions
\bea
\lambda^\prime&=& \frac{1-e^\lambda}{x}+\frac{x}{1+2\xi_{\mathrm{BS}}\tilde\sigma^2+\xi_{\mathrm{GM}}\Delta \tilde\phi^2}\Bigg\{
e^\lambda\left((\tilde m^2+\tilde\omega^2e^{-\nu}+\frac{\lambda_{\mathrm{BS}}}{2}\tilde\sigma^2)\tilde{\sigma}^2+
\Delta\frac{\tilde\phi^2}{x^2}+\frac{\lambda_{\mathrm{GM}}}{4}\Delta^2(1-\tilde\phi^2)^2\right)\nonumber\\
&&+(1+4\xi_{\mathrm{BS}})\tilde{\sigma}^{\prime 2}+\frac 12 (1+4\xi_{\mathrm{GM}})\Delta \tilde\phi^{\prime 2}-2\xi_{\mathrm{BS}}\nu^\prime \tilde{\sigma}\tilde{\sigma}^\prime-\xi_{\mathrm{GM}}\Delta\nu^\prime\tilde\phi\tilde\phi^\prime\nonumber\\
&&+4\xi_{\mathrm{BS}} e^\lambda\left[\tilde m^2-\tilde\omega^2 e^{-\nu}+\lambda_{\mathrm{BS}}\tilde{\sigma}^2-\xi_{\mathrm{BS}}\tilde R\right]\tilde{\sigma}^2\nonumber\\
&&+2\xi_{\mathrm{GM}}\Delta e^\lambda\left[\lambda_{\mathrm{GM}}\Delta(\tilde\phi^2-1)+\frac{2}{x^2}-\xi_{\mathrm{GM}}\tilde R\right]\tilde\phi^2\Bigg\},
\label{lambda BSGM}
\eea
\bea
\nu^\prime&=&\frac{x}{1+2\xi_{\mathrm{BS}} \tilde{\sigma}^2+2\xi_{\mathrm{BS}} x \tilde{\sigma}\tilde{\sigma}^\prime+\xi_{\mathrm{GM}}\Delta\tilde\phi^2+\xi_{\mathrm{GM}}\Delta x\tilde\phi\tilde\phi^\prime}\Bigg\{
\frac{-1+e^\lambda}{x^2}(1+2\xi_{\mathrm{BS}}\tilde{\sigma}^2+\xi_{\mathrm{GM}}\Delta\tilde\phi^2)\nonumber\\
&&+\tilde{\sigma}^{\prime 2}-e^\lambda(\tilde m^2-\tilde\omega^2 e^{-\nu}+\frac{\lambda_{\mathrm{BS}}}{2}\tilde{\sigma}^2)\tilde{\sigma}^2-\frac{8\xi_{\mathrm{BS}}\tilde{\sigma}\tilde{\sigma}^\prime}{x}+\Delta\frac{\tilde\phi^{\prime 2}}{2}-\frac{4\xi_{\mathrm{GM}}\Delta\tilde\phi\tilde\phi^\prime}{x}\nonumber\\
&&-\Delta e^\lambda\left[\frac{\tilde\phi^2}{x^2}+\frac{\lambda_{\mathrm{GM}}}{4}\Delta(1-\tilde\phi^2)^2\right]\Bigg\}.
\label{nu BSGM}
\eea
Instead of using the ($\theta\theta$) Einstein equation (or the equivalent ($\varphi\varphi$) equation), we use the trace equation, $G_\mu^\mu=-R=8\pi G_N T_\mu^\mu$,   leading to the rescaled Ricci scalar
\bea
\tilde R&=&\frac{2\tilde m^2\tilde{\sigma}^2+2(1+6\xi_{\mathrm{BS}})\left[(\tilde m^2-\tilde\omega^2e^{-\nu}+\lambda_{\mathrm{BS}}\tilde{\sigma}^2)\tilde{\sigma}^2+e^{-\lambda}\tilde{\sigma}^{\prime 2}\right]}
{1+2\xi_{\mathrm{BS}}(1+6\xi_{\mathrm{BS}})\tilde{\sigma}^2+\xi_{\mathrm{GM}}(1+6\xi_{\mathrm{GM}})\Delta\tilde\phi^2}\nonumber\\
&+&\Delta \frac{\lambda_{\mathrm{GM}}\Delta(1-\tilde\phi^2)+(1+6\xi_{\mathrm{GM}})\left[e^{-\lambda}\tilde\phi^{\prime 2}+\frac{2\tilde\phi^2}{x^2}-\lambda_{\mathrm{GM}}\Delta(1-\tilde\phi^2)\tilde\phi^2\right]}{1+2\xi_{\mathrm{BS}}(1+6\xi_{\mathrm{BS}})
\tilde{\sigma}^2+\xi_{\mathrm{GM}}(1+6\xi_{\mathrm{GM}})\Delta\tilde\phi^2}.
\label{Ricci BSGM}
\eea
In the limit $r\rightarrow \infty$ ($\tilde\phi\rightarrow 1$ and $\tilde\sigma\rightarrow 0$) Eqs.~(\ref{lambda BSGM}) and~(\ref{nu BSGM})  can be formally integrated yielding
\beq
e^{-\lambda(r)}=e^{\nu(r)}=1-\frac{\Delta}{1+\xi_{\mathrm{GM}}\Delta}-\frac{2 G_N  M}{r},
\label{metrika infty}
\eeq
where $M$ is an integration constant. In analogy with the space-times without deficit solid angle for which the metric function is written in terms of the mass function $g_{rr}^{-1}=1-2 G_N m(r)/r$, for  finite $r$ we have $M= M(r)$, and so we shall name $M(r)$ the \emph{core mass function}. Besides, the deficit solid angle is modified due to the presence of nonminimal coupling:
\beq
\tilde\Delta=\frac{\Delta}{1+\xi_{\mathrm{GM}}\Delta}.
\label{TildeDelta}
\eeq

Now we proceed to solving the system that consists of a boson star and a global monopole which interact only gravitationally. The set of nonlinear differential equations (\ref{eom BS BSGM}-\ref{nu BSGM}) with (\ref{Ricci BSGM})  upon providing the boundary conditions
\bea
&&\lambda(0)=0,\qquad \nu(\infty)=\ln(1-\tilde\Delta),\nonumber\\
&& \tilde{\sigma}(0)=\tilde{\sigma}_0,\qquad \tilde{\sigma}(\infty)=0,\nonumber \\
&&\tilde\phi(0)=0,\qquad\tilde\phi(\infty)=1,
\label{BC BSGM}
\eea
is solved numerically using the software code \textsc{colsys}~\cite{Colsys}.
 For this purpose we map an infinite space $x\in[0,\infty>$ to the interval $\tilde{x}\in[0,1]$ by virtue of the transformation
 $x=\tilde{x}/(1-\tilde{x})$.\\
All physical quantities involved to describe this system are given in the reduced Planck units, and are not very different from unity. Since in physically
interesting situations these parameters may wildly differ from unity, it is important to
 observe that
 Eqs.~(\ref{eom BS BSGM}--\ref{Ricci BSGM}) are invariant under the following (rescalings) transformations
\beq
x\rightarrow \beta x,\quad\tilde\sigma\rightarrow\tilde\sigma,
\quad \tilde\phi\rightarrow \tilde\phi,\quad \lambda_{\mathrm{BS,GM}} \rightarrow \frac{\lambda_{\mathrm{BS,GM}}}{\beta^2},\quad \xi_{\mathrm{BS,GM}}\rightarrow \xi_{\mathrm{BS,GM}},\quad \tilde{R} \rightarrow \frac{\tilde{R}}{\beta^2}.
\eeq
The total mass scales as
\beq
M \rightarrow \beta M.
\eeq

\section{Basic features of the nonminimal global monopole}\label{NGM}
In this section we shall only briefly discuss the basic monopole characteristics that are crucial for Sec.~\ref{BSGM}. There has been a large amount of papers discussing the gravitational field of the global monopole (see \emph{e.g.}~\cite{monGF1,monGF2,monGF3,monGF4}). Nucamendi \emph{et al}.~\cite{NucamendiBO} extended the gravitating global monopole by introducing nonminimal coupling. The main result of their analysis is the existence of bound orbits which are not present in the minimally coupled global monopole.
   However, an in-depth analysis of nonminimally coupled global monopoles as compact objects (in terms of energy density, pressures, compactness \emph{etc.}), is still lacking. In this section we bridge that gap.
\subsubsection{Deficit solid angle}
The meaning of the (modified) deficit solid angle is best understood if we transform the asymptotic  metric
\beq
ds^2=-\left(1-\tilde\Delta-\frac{2G_{\mathrm{N}} M}{r}\right)dt^2+\frac{dr^2}{1-\tilde\Delta-\frac{2G_{\mathrm{N}} M}{r}}+r^2d\Omega^2\,
\label{asymtotic GM metric}
\eeq
by virtue of  global coordinate transformations
\beq
\tilde r^2=\frac{r^2}{1-\tilde\Delta},\quad \tilde t^2=(1-\tilde\Delta)t^2,
\eeq
to the following form
\beq
ds^2=-\left(1-\frac{2\tilde G_{\mathrm{N}} \tilde M}{\tilde r}\right)d\tilde t^2+\frac{d\tilde r^2}{1-\frac{2\tilde G_{\mathrm{N}} \tilde M}{\tilde r}}+(1-\tilde\Delta) \tilde r^2d\Omega^2,
\eeq
with $\tilde{G}_N \tilde M=G_N M (1-\tilde\Delta)^{3/2}$. It is now transparent that $\tilde\Delta$ has a meaning of the deficit solid angle -- that is, the surface area of the sphere with a radius $\tilde r$ is now $4\pi(1-\tilde\Delta)\tilde r^2$. Hence, the gravitational field outside the monopole is characterized by the core mass and the solid angle deficit (proportional to the spontaneous symmetry breaking energy scale).

\subsubsection{Core mass Newtonian force}
From the asymptotic metric~(\ref{asymtotic GM metric})
which is valid only for large $r$, we can assume that the $g_{rr}$ component for smaller $r$ can be written in terms of the core mass function $M(r)$
\beq
g_{rr}^{-1}=1-\tilde\Delta-\frac{2G_{\mathrm{N}} M(r)}{r},
\eeq
from which follows the Newtonian potential generated by the core mass function
\beq
\phi_M(r)=-\frac{G_{\mathrm{N}} M(r)}{r}.
\label{NewtonV}
\eeq
A test particle with an angular momentum (per unit mass)  $L$ will feel the Newtonian force generated by the core mass function
\beq
F_M(r)=-\nabla \phi_M(r)+\frac{L^2}{r^3}.
\label{NewtonFM}
\eeq
\subsubsection{Compactness}
If we now rewrite $g_{rr}^{-1}$ in a slightly different form
\beq
g_{rr}^{-1}=(1-\tilde\Delta)\left(1-\frac{1}{1-\tilde\Delta}\frac{2 G_{\mathrm{N}} M(r)}{r}\right)
\eeq
we can read off the compactness function
\beq
\mu(r)=\frac{1}{1-\tilde\Delta}\frac{2 G_{\mathrm{N}} M(r)}{r}\,,
\label{mu}
\eeq
from which it follows that, in order to avoid event horizon formation, the compactness function must be less then unity.

\subsection{Results for $\xi_{\mathrm{GM}}=0$}\label{NGM R1}
Let us briefly demonstrate the basic monopole characteristics using a toy model that consists of the monopole
field that corresponds to the pure false vacuum inside the core and the exactly
true vacuum at the exterior~\cite{Lousto}:
\beq
\phi(r)=\Big\{\begin{array}{c} 0 \quad\mbox{if}\quad r<\delta\\
                                     \phi_0 \quad\mbox{if}\quad r>\delta,
                           \end{array}
\eeq
where $\delta$ is the so called monopole core radius. Solution to
the Einstein equations in the interior $\left(r<\delta\right)$ is the
de Sitter metric
\beq
e^{-\lambda}=e^\nu=1-H^2 r^2,
\label{de Sitter limes}
\eeq
with $H^2=2\pi G_{\mathrm{N}}\lambda_{\mathrm{GM}} \phi_0^4/3$. The metric outside the monopole core $\left(r>\delta\right)$
is the asymptotic metric (\ref{asymtotic GM metric}). The core radius $\delta$ and the core mass $M$ are determined by continuously matching the interior and exterior
metrics yielding
\beq
\delta=\frac{2}{\sqrt{\lambda_{\mathrm{GM}}}\phi_0},\qquad M=-\frac{16\pi }{3}\frac{\phi_0}{\sqrt{\lambda_{\mathrm{GM}}}}.
\label{CoreSize}
\eeq
This toy model result is very interesting as it shows that the monopole core mass is negative.

In the case of minimal coupling it has been shown~\cite{LieblingMon1,LieblingMon2}  that  for monotonically increasing scalar field, the regular solutions without horizon exist only for $\Delta<1$. For $1<\Delta<3$ there are regular solutions with the horizon. However, for $\Delta>3$ there are no regular solutions which has been shown to be in accord with the topological inflation (see \emph{e.g.}~\cite{LieblingMon1}). The existence of the horizon can be seen from the asymptotic behaviour of the metric functions: in the limit $r\rightarrow\infty$ the metric functions $g_{tt}$ and $g_{rr}^{-1}$ approach zero for $\Delta\rightarrow 1$. In this paper we are interested in the regular monopole solutions without horizon.
\begin{figure}
\centering
\includegraphics[scale=0.75]{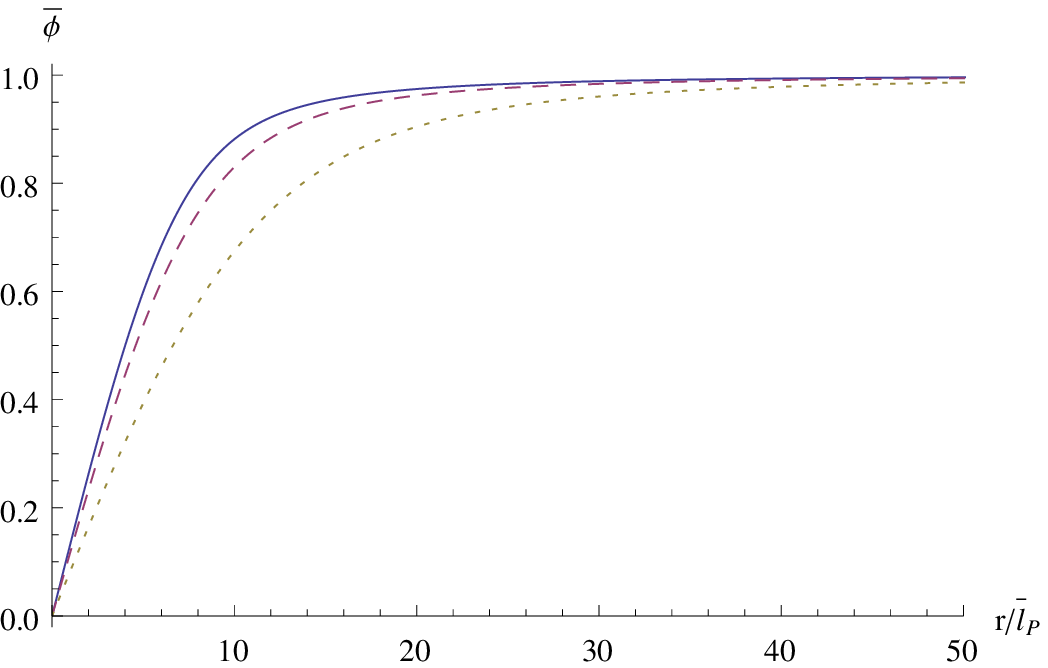}\quad
\includegraphics[scale=0.75]{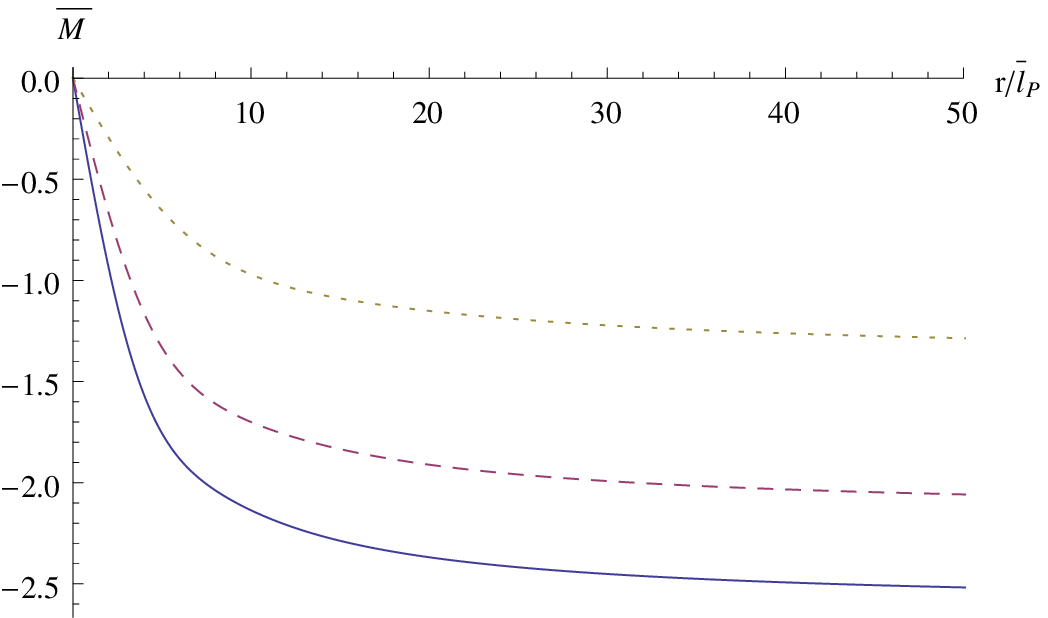}\\
\caption{The monopole field $\bar\phi$ (left plot) and the core  mass functions (right plot) for $\lambda_{\mathrm{GM}}=0.1$, $\xi_{\mathrm{GM}}=0$, $\Delta=0.999$ for solid, $\Delta=0.7$ for dashed and $\Delta=0.3$ for dotted curves.}
\label{SigmeMetrikeGM}
\end{figure}%

\begin{figure}
\centering
\includegraphics[scale=0.5]{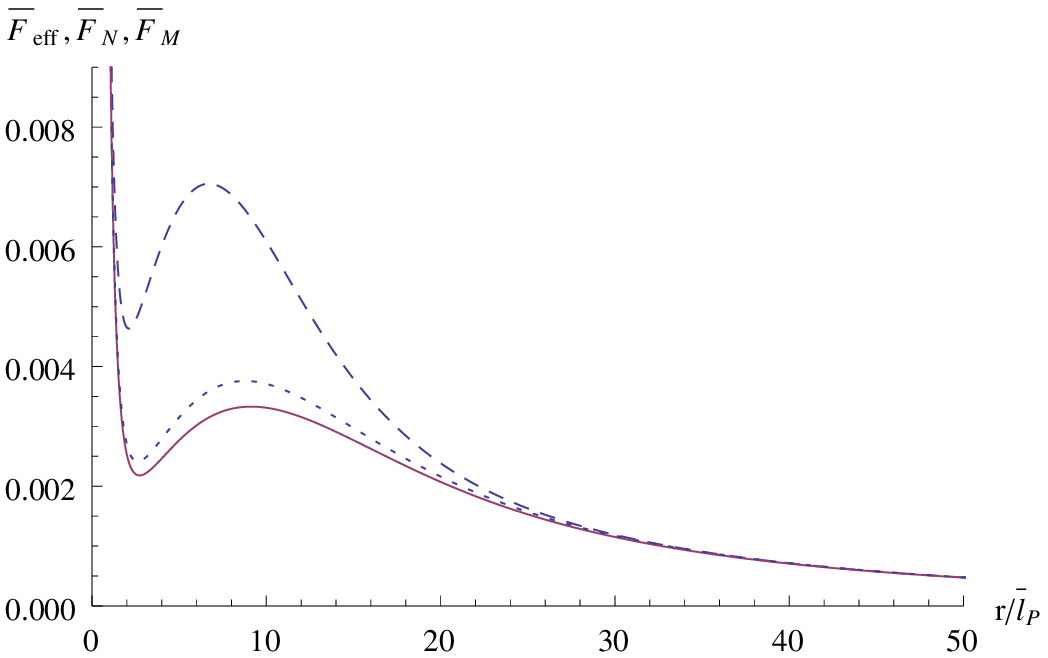}\quad
\includegraphics[scale=0.5]{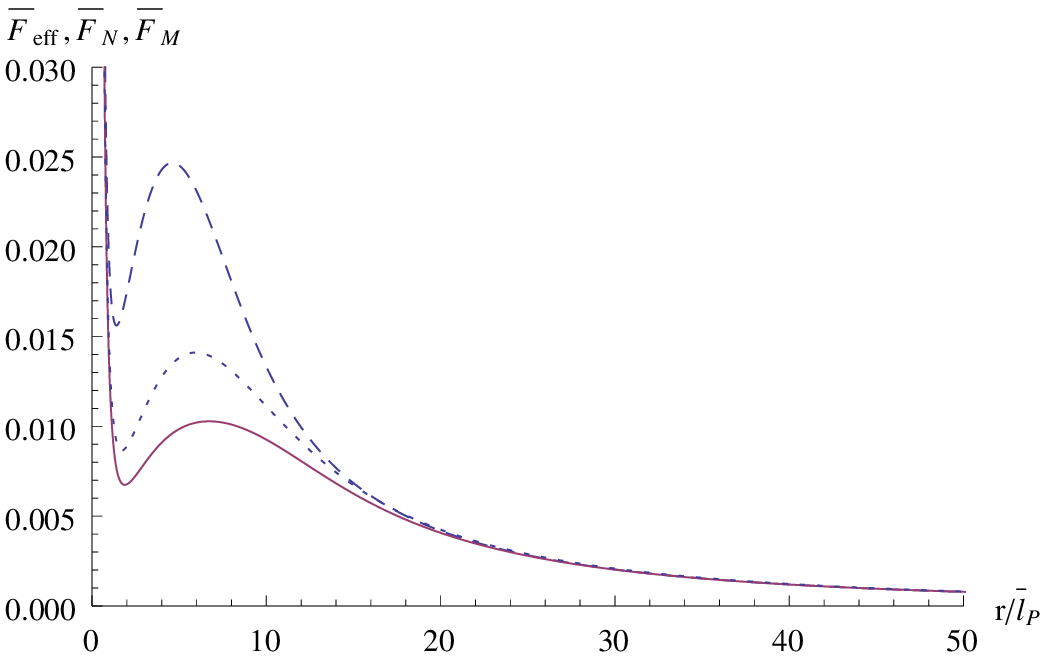}\quad
\includegraphics[scale=0.5]{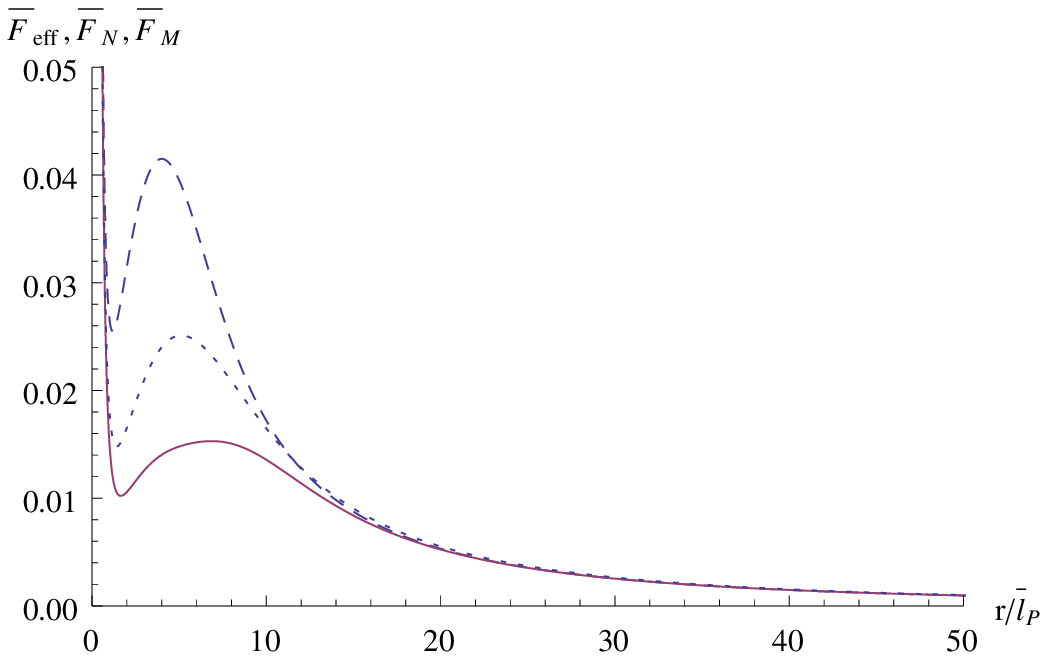}
\caption{The effective $\bar F_{\rm eff}$ forces~(\ref{Feff}) (solid curves), the Newtonian forces $\bar F_N$~(\ref{NewtonF}) (dotted curves) and the Newtonian forces produces by the core mass functions~(\ref{NewtonFM}) (dashed curves) for $\Delta=0.3$ (left plot), $\Delta=0.7$ (middle plot) and $\Delta=0.999$ (right plot). Also the angular momentum and the energy per unit mass are $\bar L=0.1$ and  $\bar E=1$.}
\label{SileGM}
\end{figure}

 First, it is instructive to explore the effect of different symmetry breaking scales $\Delta$ on the monopole configuration in the minimal coupling case. Hence in Figs.~\ref{SigmeMetrikeGM}-\ref{SileGM} we plot all relevant functions of the monopole configurations for three different values of $\Delta$: $\Delta=0.3$ for dotted, $\Delta=0.7$ for dashed and $\Delta=0.999$ for solid curves. In the left plot of Fig.~\ref{SigmeMetrikeGM} we see that the change of $\Delta$ only slightly influences the shape of the monopole field - the fields remain monotonic.
The core mass function is more negative for larger $\Delta$ as seen in the right plot of Fig.~\ref{SigmeMetrikeGM}. Correspondingly, the Newtonian forces produced by the core masses~(\ref{NewtonFM}) (dashed curves), the Newtonian forces~(defined in Appendix A, Eq.~(\ref{NewtonF})) (dotted curves) and also the effective forces~(defined i n Appendix B, Eq.~(\ref{Feff})) (solid curves) are more repulsive for larger $\Delta$ as shown in Fig.~\ref{SileGM}. For small $\Delta$ the Newtonian force is in agreement with the effective force as seen in the left and middle plots of Fig.~\ref{SileGM}. Since the effective force does not cross zero, there exists no bound orbits for the minimally coupled monopole. This result is  interesting and in a way represents a signature of the global monopole configuration in the minimal setting -- even though the energy density is positive and decreasing as $1/r^2$, a particle moving in a monopole field feels a repulsive force.
  This is due to the fact that the total mass, which is obtained as the volume integral of the energy density, can be written as a sum of two parts: one part comes from the core mass function and the other part comes from the deficit solid angle. Only the core mass function contributes to the Newtonian force as the second part is linear in $r$, thus yielding a constant Newtonian potential which produces no Newtonian force. In all three cases both Newtonian forces are in qualitative agreement with the effective forces for small $r$ while for large $r$ all forces agree very well both qualitatively and quantitatively, as expected. The differences between the Newtonian and the effective forces can be traced back to the \emph{gravitational slip} which is defined as the difference between the two Newtonian potentials (the one corresponds to $g_{tt}$ and the other to $g_{rr}$) and which is known to be different from zero in the presence of matter.
\begin{figure}
\centering
\includegraphics[scale=0.8]{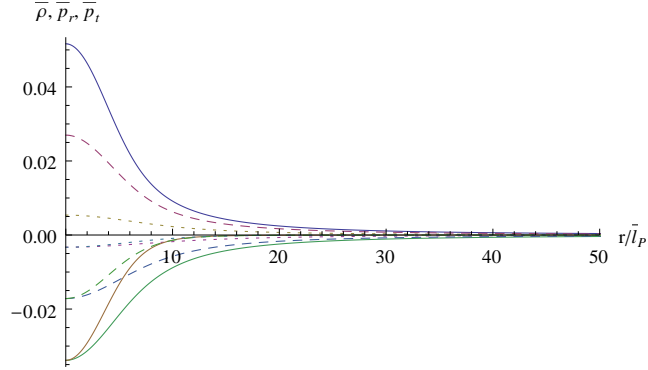}
\caption{The energy density (positive definite), the radial and the tangential pressures (negative-definite with $p_t>p_r$)  for $\lambda_{\mathrm{GM}}=0.1$, $\xi_{\mathrm{GM}}=0$, $\Delta=0.999$ for solid, $\Delta=0.7$ for dashed and  $\Delta=0.3$ for dotted curves.}
\label{MaseSileGM}
\end{figure}
Independently on the value of $\Delta$, the energy density is always positive-definite while the pressures are negative-definite functions of the radial coordinate in the minimal coupling case. Also the magnitude of all three thermodynamic functions are larger for larger $\Delta$ as shown in Fig.~\ref{MaseSileGM}. In all these cases the strong energy condition is violated and this trend is more prominent for larger $\Delta$.

\subsection{Results for $\xi_{\rm{GM}}\ne0$}\label{NGM R2}

\begin{figure}
\centering
\includegraphics[scale=0.8]{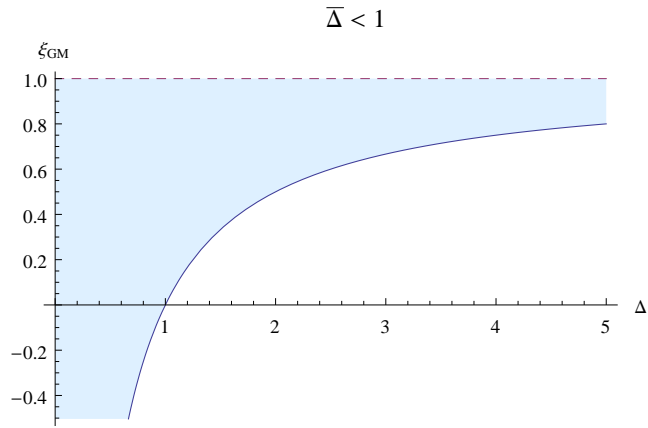}
\caption{The nonminimal coupling $\xi_{\mathrm{GM}}$ as a function of the symmetry breaking scale $\Delta$ for $\tilde\Delta=1$. The shaded region shows allowed values of $\xi_{\mathrm{GM}}$ and $\Delta$ for $\tilde\Delta<1$.}
\label{figDelta}
\end{figure}

In this subsection we show how nonminimal coupling of the monopole field to gravity affects the behaviour of minimally coupled global monopole studied in the previous subsection. This analysis was firstly obtained by Nucamendi \emph{et al}.~\cite{NucamendiBO} and our results are in agreement with theirs. Just like in the case of boson stars, nonminimal coupling drastically changes the monopole configuration. Firstly, from the asymptotic behaviour of the metric function~(\ref{asymtotic GM metric}) we observe that an event horizon forms if $1-\tilde\Delta-2G_N M(r)/r=0$ for a finite $r$, where $\tilde\Delta$ is given in Eq.~(\ref{TildeDelta}). In this paper we shall not consider configurations with an event horizon and hence we shall demand that $2G_N M(r)/r<1-\tilde\Delta$ for all finite $r$. When $r\rightarrow\infty$, $G_N M(r)/r\rightarrow 0$ and the above condition reduces to $\tilde\Delta< 1$.
In Fig.~\ref{figDelta} we show how $\xi_{\mathrm{GM}}$ depends on $\Delta$ if we demand that $\tilde\Delta<1$.
The shaded region shows allowed values of $\xi_{\mathrm{GM}}$ for a given $\Delta$.
Here, for example, we have the situation that  the horizon will not form for $\Delta=1$ if $\xi_{\mathrm{GM}}$ is only slightly greater then zero. Besides, if $\xi_{\mathrm{GM}}\ge 1$ there are no restrictions on $\Delta$. This result is very important as it clearly shows that in the nonminimal case, there are much more allowed values for the energy of the symmetry breaking scale $\Delta$ that lead to regular solutions without horizons.

\begin{figure}
\centering
\includegraphics[scale=0.75]{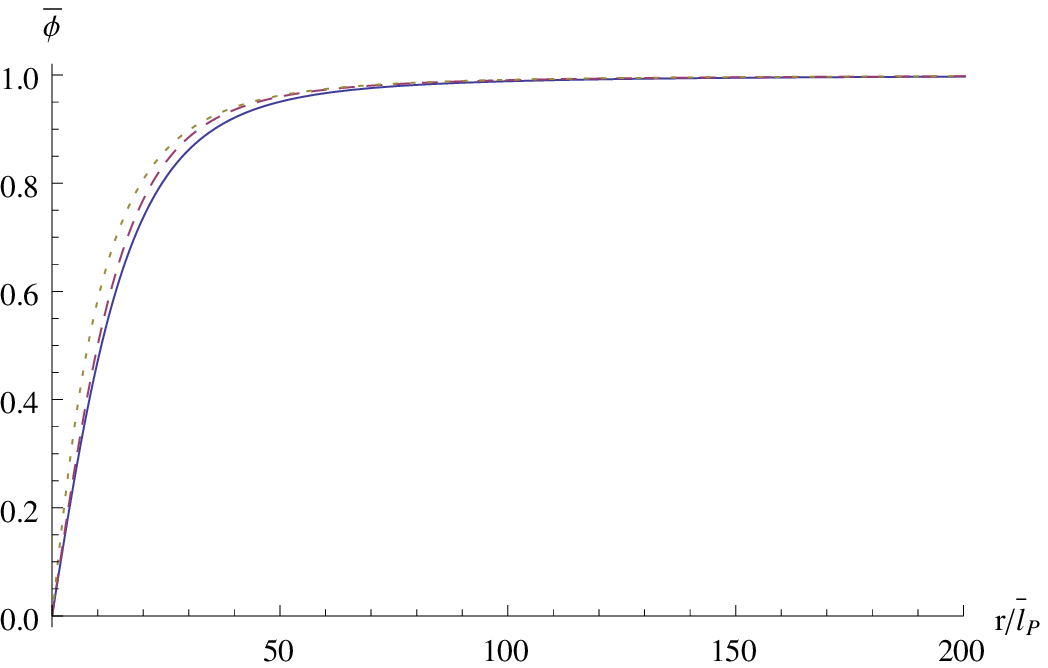}\quad
\includegraphics[scale=0.75]{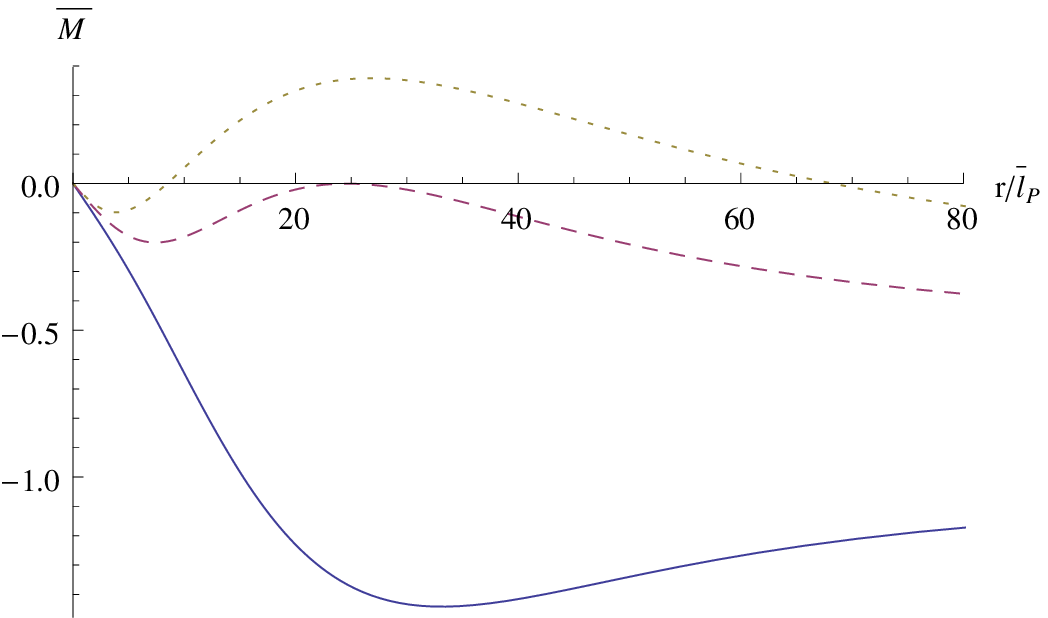}
\caption{The monopole field $\bar\phi$ in the left plot and the core mass functions in the right plot for $\lambda_{\mathrm{GM}}=0.1$, $\Delta=0.1$, $\xi_{\mathrm{GM}}=-1$ for solid, $\xi_{\mathrm{GM}}=1$ for dashed and $\xi_{\mathrm{GM}}=2$ for dotted curves.}
\label{SigmeMetrikeNGM}
\end{figure}

\begin{figure}
\centering
\includegraphics[scale=0.5]{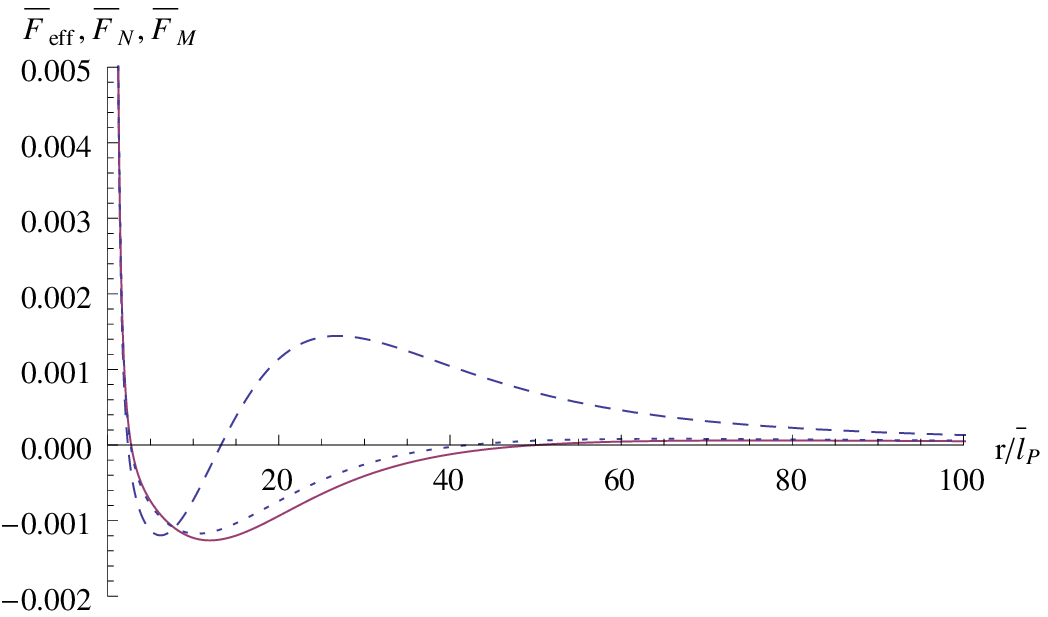}\quad
\includegraphics[scale=0.5]{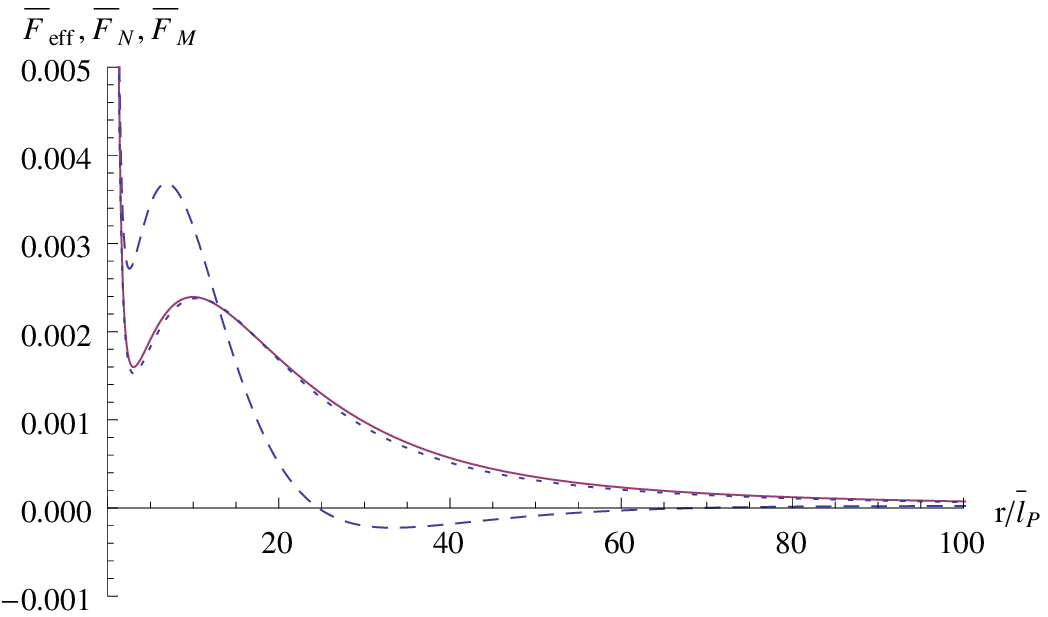}\quad
\includegraphics[scale=0.5]{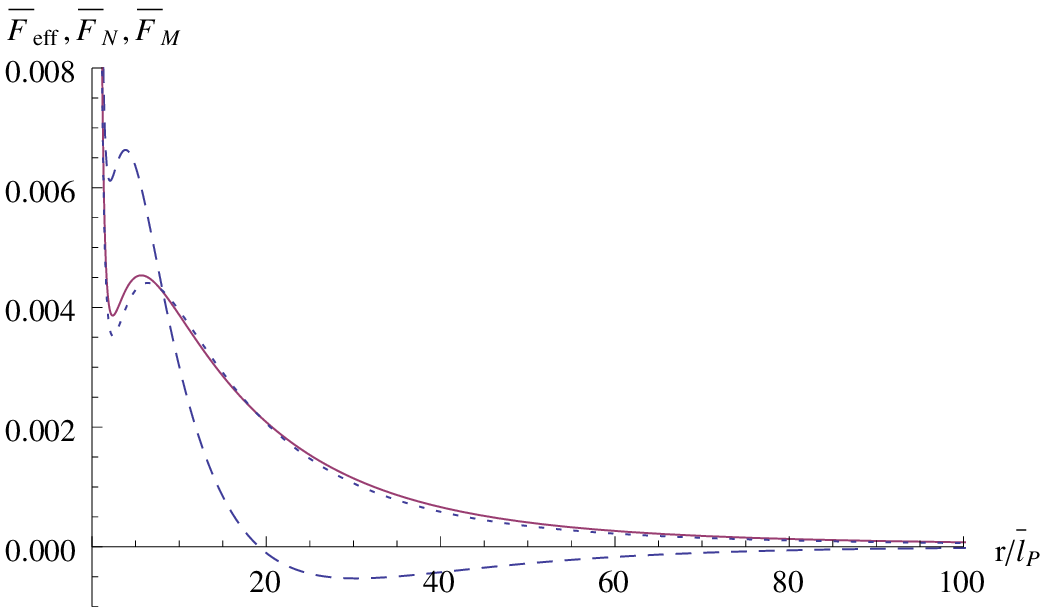}
\caption{The effective $\bar F_{\rm eff}$ forces~(\ref{Feff}) (solid curves), the Newtonian forces $\bar F_N$~(\ref{NewtonF}) (dotted curves) and the Newtonian forces produces by the core mass functions~(\ref{NewtonFM}) (dashed curves) for $\xi_{\mathrm{GM}}=-1$ in the left, $\xi_{\mathrm{GM}}=1$ in the middle and $\xi_{\mathrm{GM}}=2$ in the right plot. The other parameters are $\lambda_{\mathrm{GM}}=0.1$, $\Delta=0.1$. Also the angular momentum and energy (of the particle) per unit mass are $\bar L=0.1$ and $\bar E=1$.}
\label{SileNGM}
\end{figure}

The most important feature of nonminimal global monopole is the existence of bound orbits which can be traced back to the minima of the effective potential. In this subsection we present all relevant functions for three different values of $\xi_{\mathrm{GM}}$: $\xi_{\mathrm{GM}}=-1$ for solid, $\xi_{\mathrm{GM}}=1$ for dashed and $\xi_{\mathrm{GM}}=2$ for dotted curves. Here the energy of the symmetry breaking scale is fixed and equals $\Delta=0.1$. The self-coupling is also fixed $\lambda_{\mathrm{GM}}=0.1$.
Even though the effect of nonminimal coupling is large, the monopole field still retains its monotonic behaviour as seen in the left plot of Fig.~\ref{SigmeMetrikeNGM}.
The core mass function is not so immune to the $\xi_{\mathrm{GM}}$-parameter as shown in the right plot of Fig.~\ref{SigmeMetrikeNGM}. For positive $\xi_{\mathrm{GM}}$, in particular for $\xi_{\mathrm{GM}}\gtrsim 1$, the core mass function as a function of the radial coordinate exhibits locally positive values.
 This trend is accompanied by the attractive force implied by the core mass function~(\ref{NewtonFM}) as shown in the middle and right plot of Fig.~\ref{SileNGM} (dashed curves). One could naively conclude that the locally positive values of the core mass functions are responsible for the existence of bound orbits.
 However, this is not the case as the effective forces~(\ref{Feff}) (solid curves) are repulsive  and they are responsible for the existence of bound orbits. While for large $r$ all three forces agree, they show significant disagreements for small $r$ inside the monopole core.

 This is an interesting result as it allows to investigate the effects of the backreaction of geometry on matter as well as how matter affects
 geometry through gravitational slip. While $ F_N$ includes the backreaction of
 geometry on matter, it does not contain nonlinear effects of matter on geometry, $F_{\rm eff}$ includes both the backreaction of geometry on matter as well as nonlinear gravitational effects.
 Finally, $F_M$ includes nonlinear effects of the geometry, but it is insensitive to the effects of
 gravitational slip. From Fig.~\ref{SileNGM}  we see that $F_N$ and $F_{\rm eff}$ agree in all cases considered, which means that the nonlinear geometrical effects are weak. In all cases $F_M$ shows qualitative disagreement with $F_{\rm eff}$ which implies that in all cases the effects of gravitational slip are significant. From the above analysis it follows that while the effective and the Newtonian force contain information of bound orbits the force produced by the core mass function does not. This then implies that the core mass function cannot be used to study bound orbits, while the \emph{active} gravitational mass (obtained by integrating $\rho+\sum p_i$ over the volume up to some radius $r$) can be used to study bound orbits.

\begin{figure}
\centering
\includegraphics[scale=0.75]{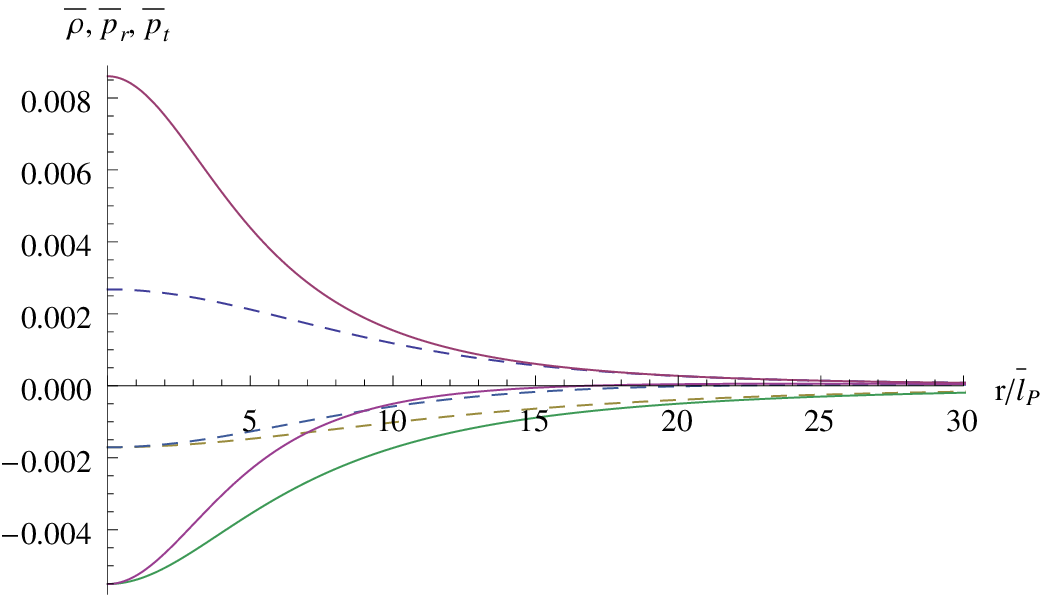}\quad
\includegraphics[scale=0.75]{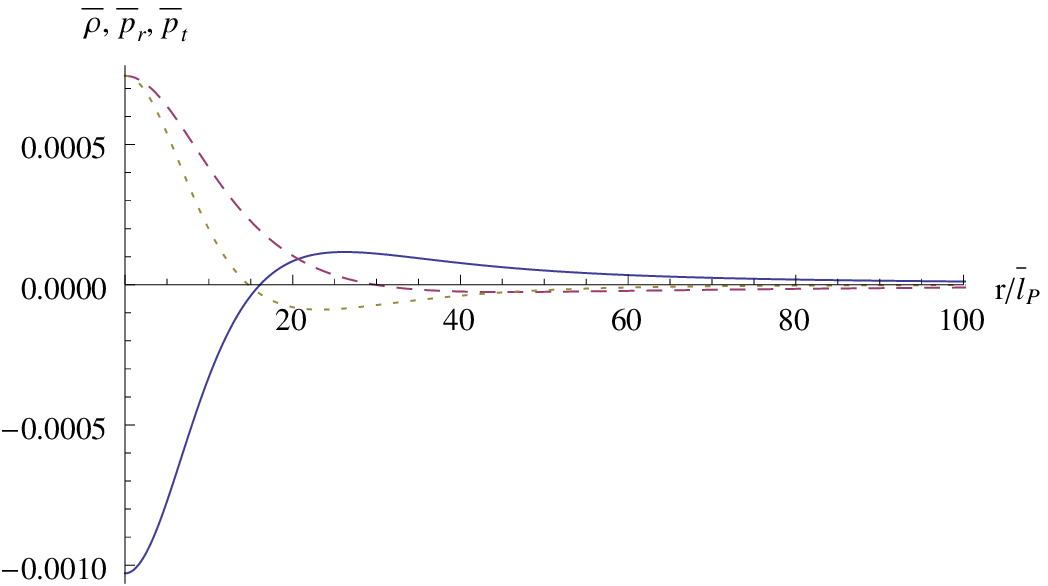}
\caption{The energy density $\bar\rho$ and the principal pressures $\bar p_r,\bar p_t$ for $\lambda_{\mathrm{GM}}=0.1$, $\Delta=0.1$ and $\xi_{\mathrm{GM}}=2$ (solid curves) and $\xi_{\mathrm{GM}}=1$ (dashed curves) in the left plot and  $\xi_{\mathrm{GM}}=-1$ in the right plot (solid curve for the energy density, dashed for the radial and dotted for the transversal pressure).}
\label{TdNGM}
\end{figure}

 It is also interesting to show how the thermodynamic functions behave for these three values of $\xi_{\mathrm{GM}}$. In the left plot of Fig.~\ref{TdNGM} we see that for positive $\xi_{\mathrm{GM}}$ the energy density is positive-definite while the pressures are negative-definite functions of the radial coordinate. However, for negative $\xi_{\mathrm{GM}}$ the energy density evolves from the negative center, crosses zero and asymptotically converge to zero from positive values while the pressures exhibit the opposite trend. So, even though the energy density is (locally) negative (and also the core mass function), the active gravitational mass is (locally) positive resulting in the (locally) attractive effective force (see left plot in Fig.~\ref{SileNGM}), implying the existence of bound orbits.
%

\section{The nonminimal boson D-star}\label{BSGM}

So far we have witnessed monotonic behaviour of both fields, the boson star (see Ref.~\cite{moiDES}) and the global monopole field (see previous section), for all choices of parameters. In the combined system this is not the case anymore: both fields reconfigure themselves depending on the parameters. Therefore, we distinguish three regimes according to the qualitative behaviour of the fields configurations:\\
\begin{itemize}
\item Weak coupling regime: in this regime both, the boson star and the global monopole fields configurations retain theirs monotonicity.
\item Mild coupling regime: in this regime the boson star field is slightly non-monotonic while the monopole field is still monotonic.
\item Strong coupling regime: in this regime both, the boson star field and the monopole field are significantly reconfigured into non-monotonic fields. This regime is particularly interesting since the boson star gets very compressed by the monopole and a whole system can reach large compactness suggesting that this object can provide a good black hole mimicker.
\end{itemize}
In Ref.~\cite{moiDES} we have seen that boson stars exhibit largest compactness for negative values of nonminimal coupling $\xi_{\rm BS}$. Here we want to analyze how the presence of a global monopole affects the boson star configuration. Therefore, we take the boson star parameters from Ref.~\cite{moiDES} with $\tilde m=1$, $\tilde{\sigma}_0=0.05$, $\lambda_{\rm BS}=0$ and $\xi_{\rm BS}=-4$ that produces an attractive effective force and combine it with the global monopole with varying parameters.
\subsection{Weak coupling regime}\label{Weak}
\begin{figure}
\centering
\includegraphics[scale=0.75]{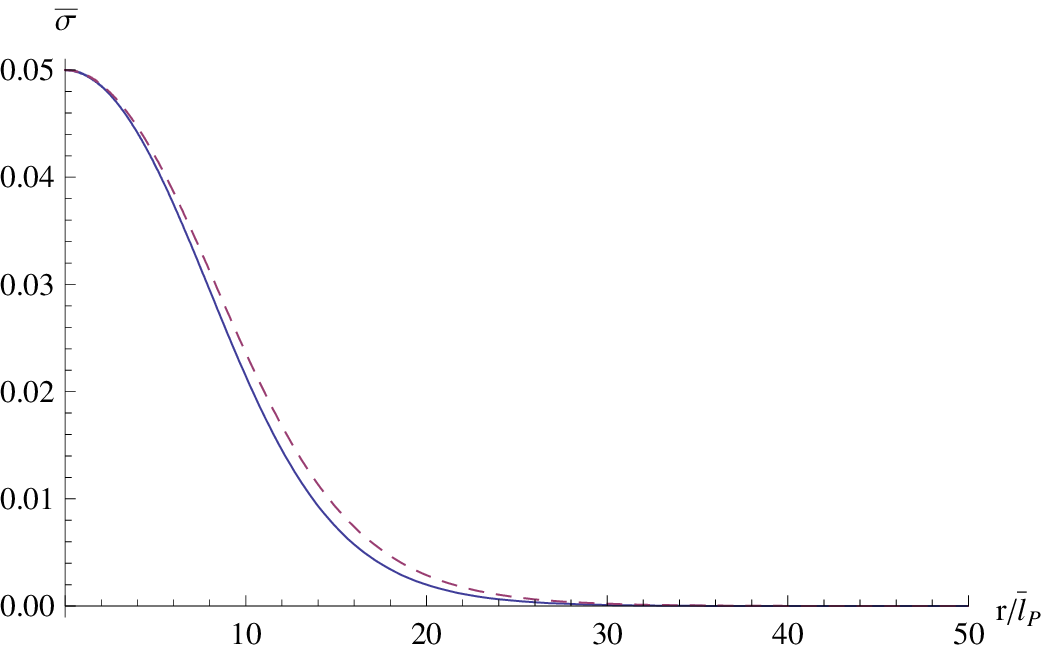}\quad
\includegraphics[scale=0.75]{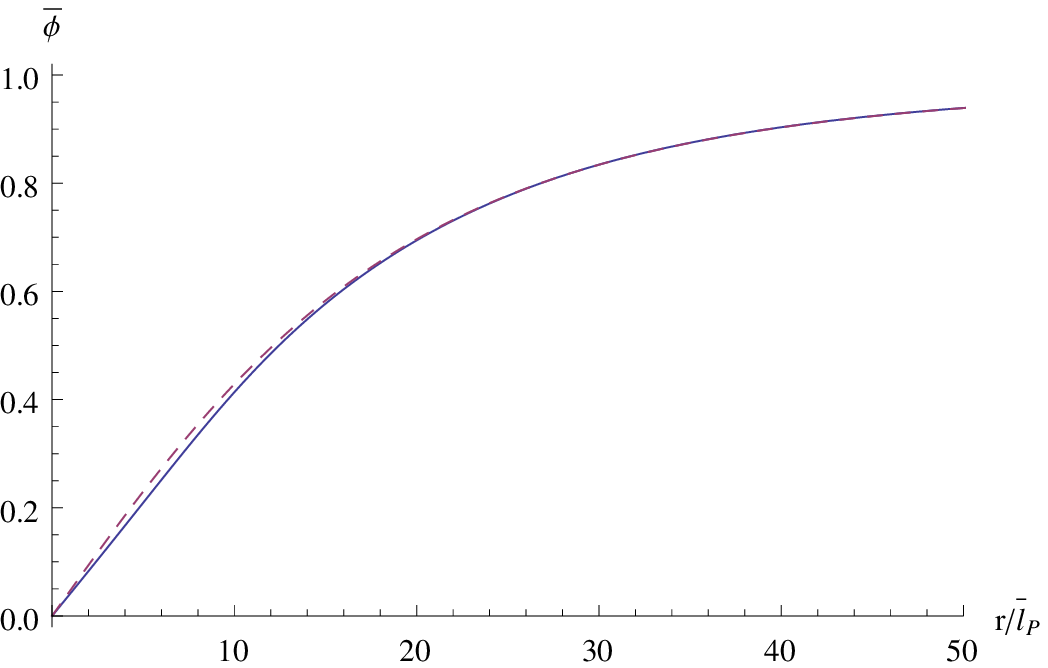}
\caption{Left plot: the boson star field in the presence of the global monopole (solid) and the boson star field alone (dashed). Right plot: the global monopole field in the presence of the boson star (solid) and the global monopole field alone (dashed). The parameters are: $\tilde{\sigma}_0=0.05$, $\lambda_{\rm BS}=0$, $\xi_{\rm BS}=-4$, $\lambda_{\rm GM}=0.1$, $\Delta=0.08$, $\xi_{\rm GM}=-1$.}
\label{poljeBSGMe}
\end{figure}
\begin{figure}
\centering
\includegraphics[scale=0.75]{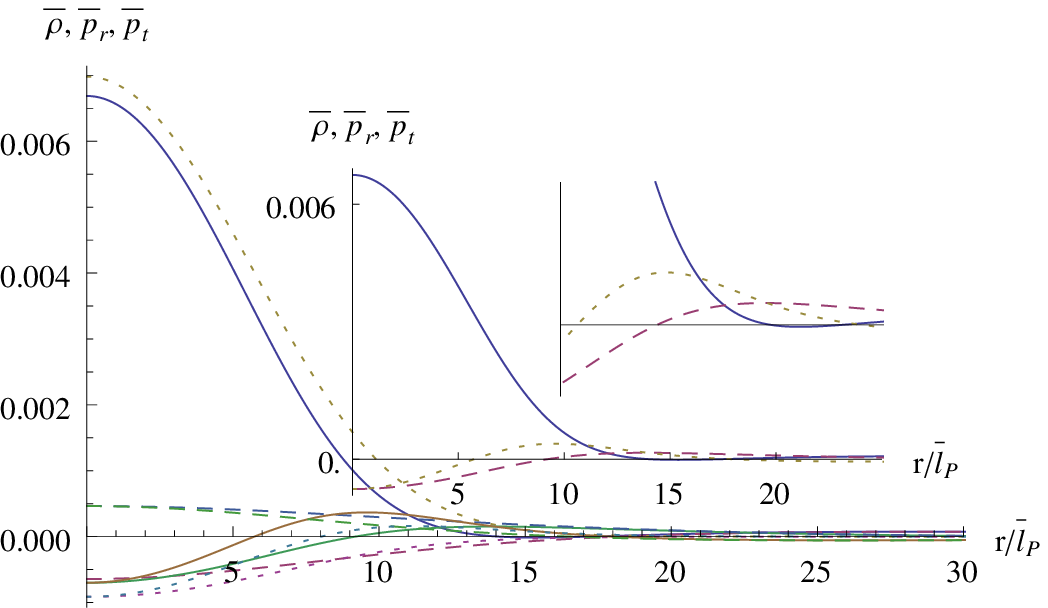}\quad
\includegraphics[scale=0.75]{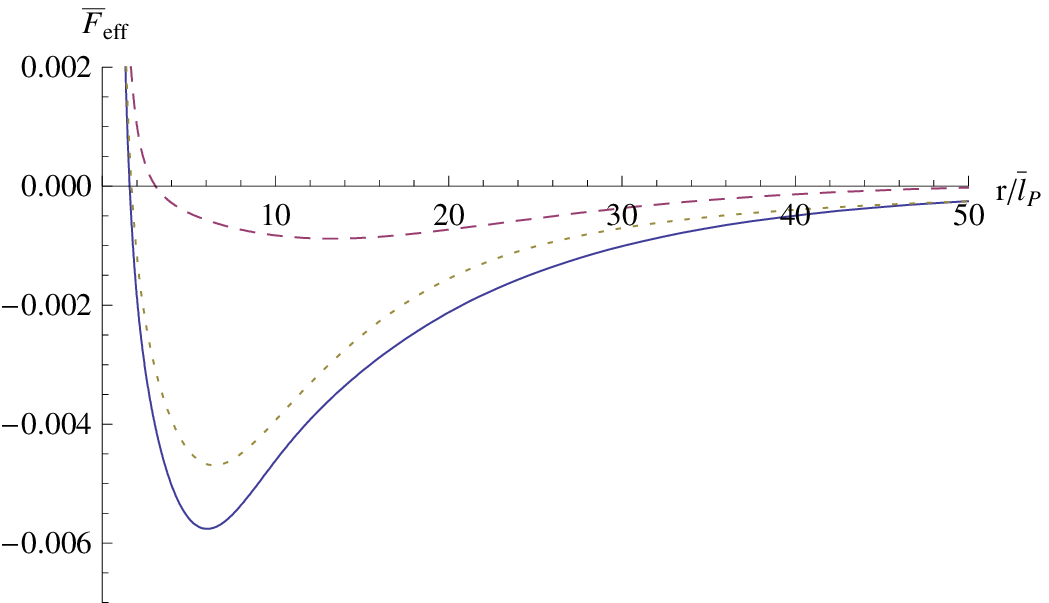}
\caption{Left plot: the energy density and the pressures for the boson star alone (dotted curves), the global monopole alone (dashed curves) and the combined system of the boson star and the global monopole (solid curves). Right plot: the effective force~(\ref{Feff}) for the boson star alone (dotted curve), the global monopole (dashed curve) and the combined system of the boson star and the global monopole (solid curve). The parameters are: $\tilde{\sigma}_0=0.05$, $\lambda_{\rm BS}=0$, $\xi_{\rm BS}=-4$, $\lambda_{\rm GM}=0.1$, $\Delta=0.08$, $\xi_{\rm GM}=-1$.}
\label{MuTdBSGMe}
\end{figure}

In this regime we take a global monopole configuration that produces an attractive effective potential, for example $\Delta=0.08$, $\lambda_{\rm GM}=0.1$ and $\xi_{\rm GM}=-1$. As seen in Fig.~\ref{poljeBSGMe} both fields, the boson star (dashed curve in the left plot) and the monopole field (dashed curve in the right plot) are only slightly affected in the combined system (solid curves). Nevertheless, it is important to point out that both fields in the combined system are reconfigured to slightly lower magnitudes, which is not the case with the combined system with the monopole that produces a repulsive effective potential.%
\begin{figure}
\centering
\includegraphics[scale=0.75]{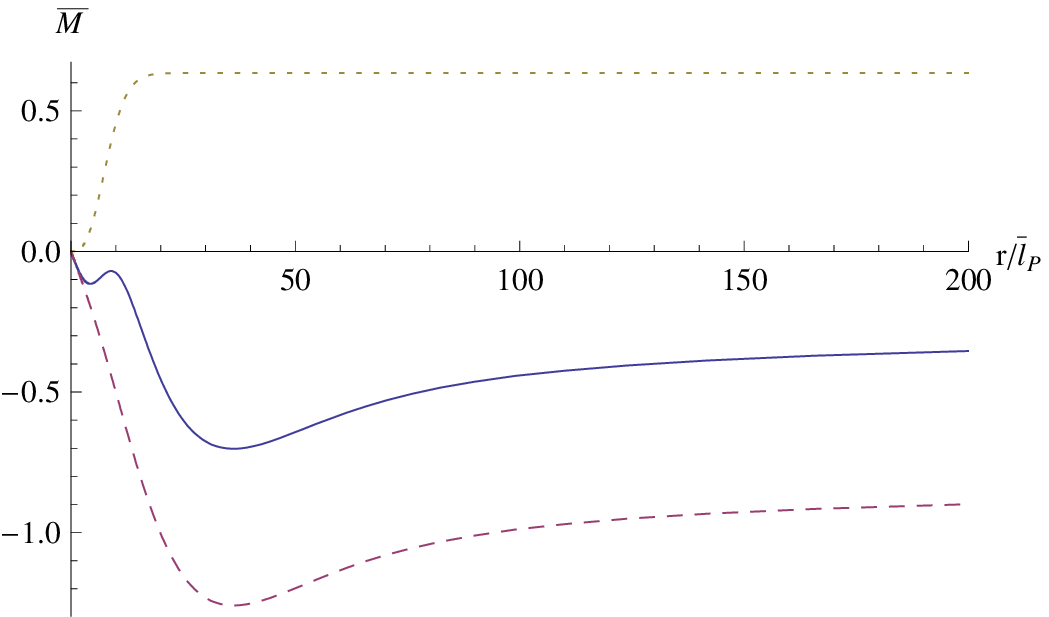}\quad
\includegraphics[scale=0.75]{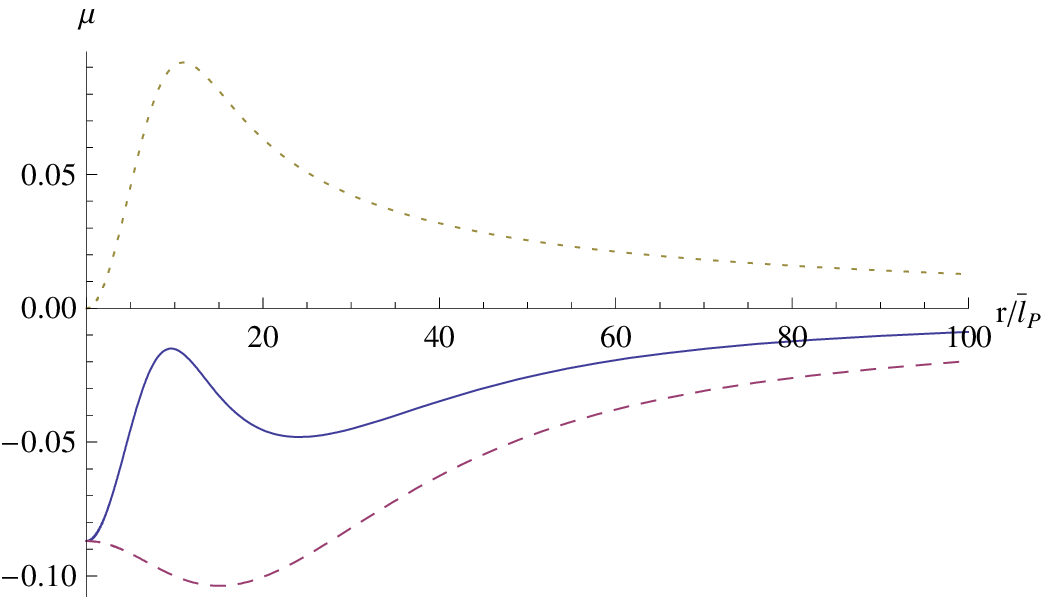}
\caption{The (core) mass function (left plot) and the compactness function (right plot) for the boson star alone (dotted curve), the global monopole alone (dashed curve) and the combined system of the boson star and the global monopole (solid curve). The parameters are: $\tilde{\sigma}_0=0.05$, $\lambda_{\rm BS}=0$, $\xi_{\rm BS}=-4$, $\lambda_{\rm GM}=0.1$, $\Delta=0.08$, $\xi_{\rm GM}=-1$.}
\label{MetrikeMaseBSGMe}
\end{figure}
The core mass function/compactness of the combined system is roughly equal to the sum of the constituent masses/compactnesses as seen in  Fig.~\ref{MetrikeMaseBSGMe}. Furthermore, the energy density and pressures in the combined system sum up approximately linearly.  Due to this fact, one can actually obtain gravastar-like pressures in the combined object
(see left plot in Fig.~\ref{MuTdBSGMe}): both pressures evolve from a negative center and exhibit a locally positive maximum, just like the gravastar pressures in their atmosphere (see, \emph{e.g.}, Refs.~\cite{MM}-\cite{Horvat:Radial}). However, the dominant energy condition is clearly violated in the combined system for the chosen set of parameters.
In the right plot of Fig.~\ref{MuTdBSGMe} we see that the effective force of the combined system (solid curve) is also approximately equal to the sum of the effective forces produced by the boson star (dotted curve) and the global monopole (dashed curve).

\subsection{Mild coupling regime}\label{Mild}
\begin{figure}
\centering
\includegraphics[scale=0.75]{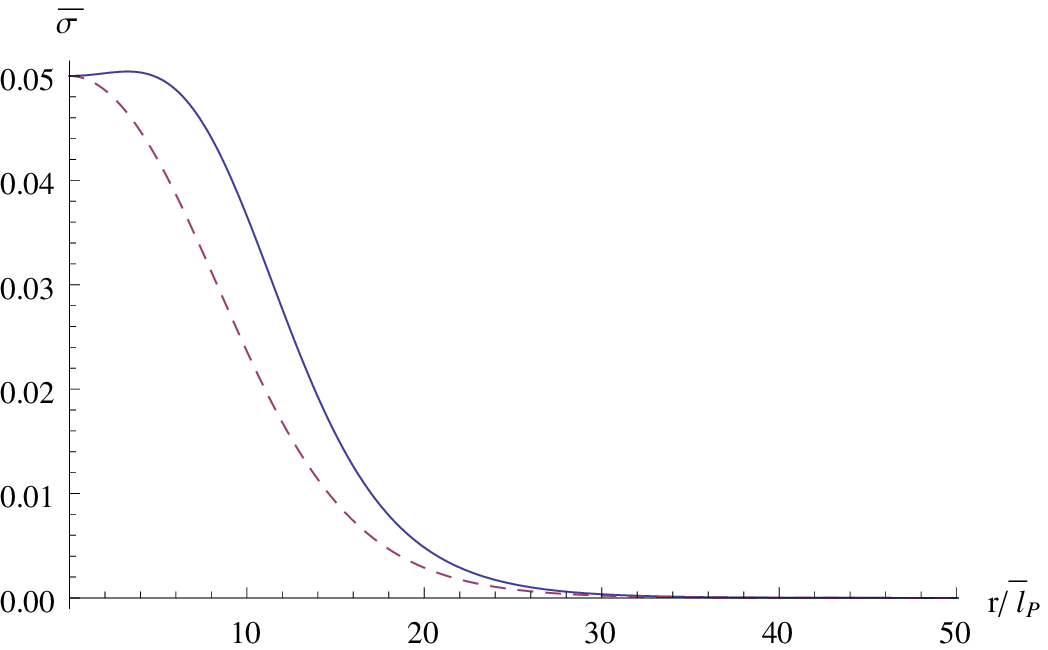}\quad
\includegraphics[scale=0.75]{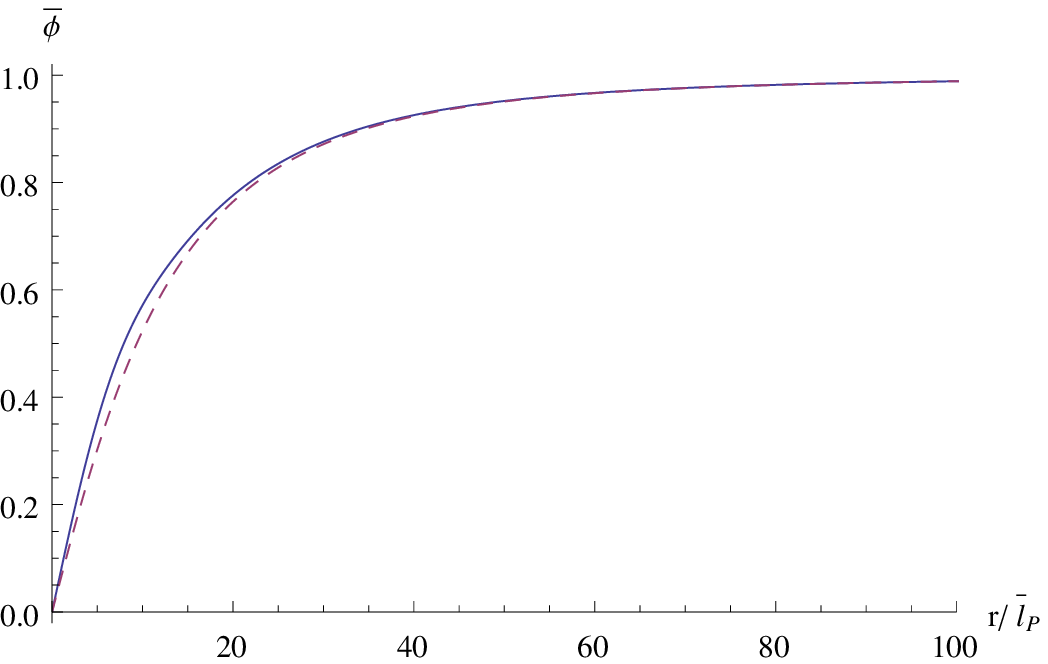}
\caption{Left plot: the boson star field in the presence of the global monopole (solid) and the boson star field in the absence of the monopole (dashed). Right plot: the global monopole field in the presence of the boson star (solid) and the global monopole field in the absence of the boson star (dashed). The parameters are: $\tilde{\sigma}_0=0.05$, $\lambda_{\rm BS}=0$, $\xi_{\rm BS}=-4$, $\lambda_{\rm GM}=0.1$, $\Delta=0.08$, $\xi_{\rm GM}=2$.}
\label{poljeBSGMa}
\end{figure}

\begin{figure}
\centering
\includegraphics[scale=0.75]{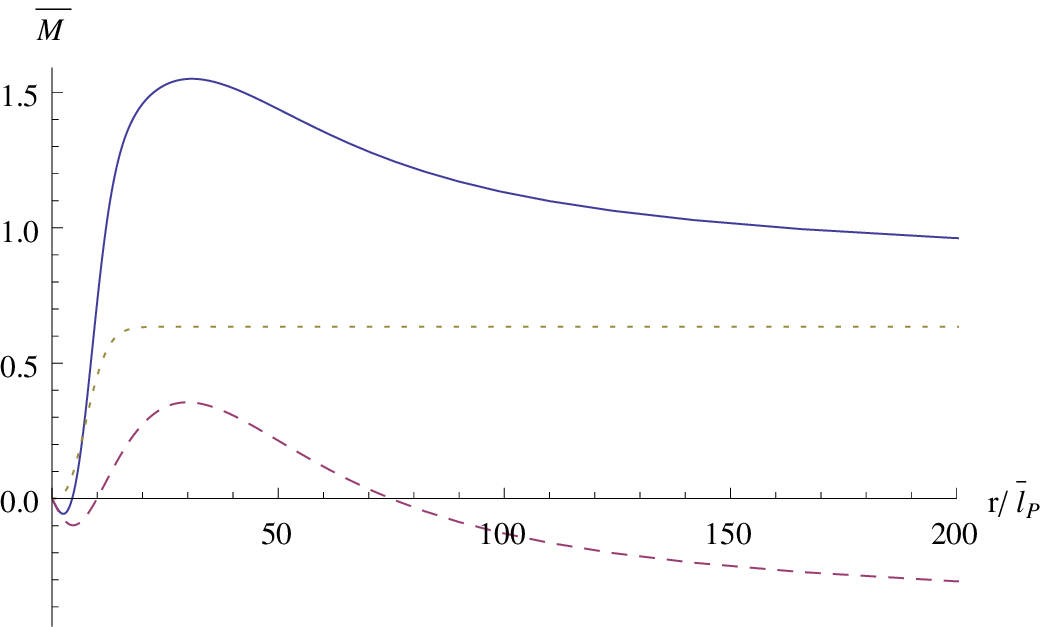}\quad
\includegraphics[scale=0.75]{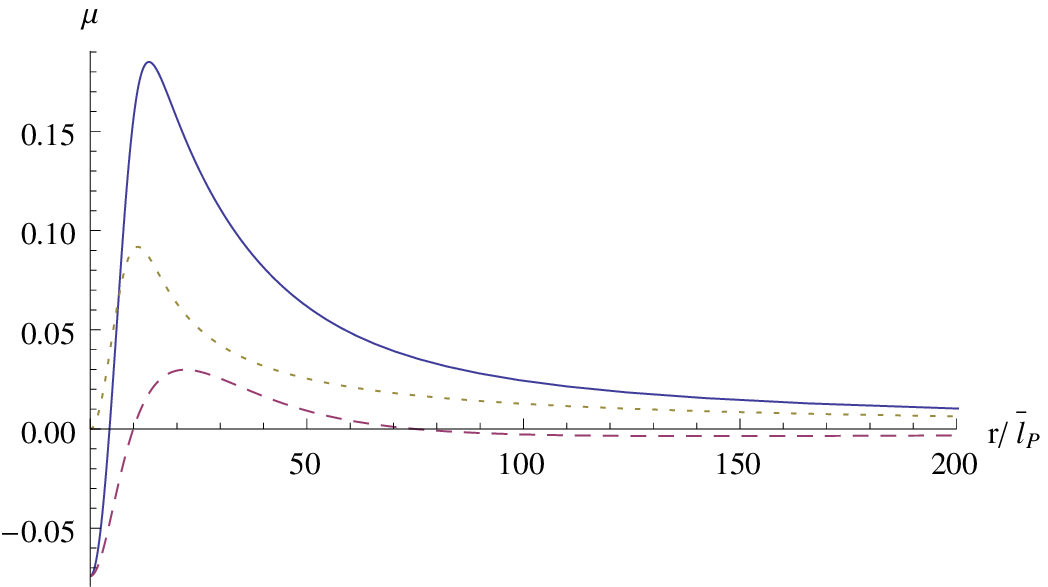}
\caption{The (core) mass function (left plot) and the compactness function (right plot) for the boson star alone (dotted curve), the global monopole alone (dashed curve) and the combined system of the boson star and the global monopole (solid curve). The parameters are: $\tilde{\sigma}_0=0.05$, $\lambda_{\rm BS}=0$, $\xi_{\rm BS}=-4$, $\lambda_{\rm GM}=0.1$, $\Delta=0.08$, $\xi_{\rm GM}=2$.}
\label{MetrikeMaseBSGMa}
\end{figure}

In this subsection we take the global monopole with $\Delta=0.08$, $\lambda_{\rm GM}=0.1$, $\xi_{\rm GM}=2$ that produces repulsive effective force.
In Fig.~\ref{poljeBSGMa} we show  $i$) in the left plot how the boson star field (dashed curve) is affected by the presence of the global monopole (solid curve) and $ii$) in the right plot how the global monopole field (dashed curve) is affected by the presence of the boson star field (solid curve). For the given set of parameters the boson star field is more sensitive to the presence of the global monopole field then vice versa: the boson star field reconfigures significantly by loosing its monotonicity while the monopole field remains monotonous.
In the left plot of Fig.~\ref{MetrikeMaseBSGMa} we show a rather unexpected behaviour of the core mass function of the combined system (solid curve), which is clearly greater then the sum of the constituents (core) masses of the boson star (dotted curve) and the global monopole (dashed curve).
This trend is important for the compactness, a function that measures how much mass can be accommodated in a certain radius. It turns out that the compactness  is significantly greater in the combined system (solid curve in the right plot of Fig.~\ref{MetrikeMaseBSGMa}) then in the boson star alone (dotted curve in the right plot of Fig.~\ref{MetrikeMaseBSGMa}) or the global monopole alone (dashed curve in the right plot of Fig.~\ref{MetrikeMaseBSGMa}), or even larger than the sum of the two.

\begin{figure}
\centering
\includegraphics[scale=0.75]{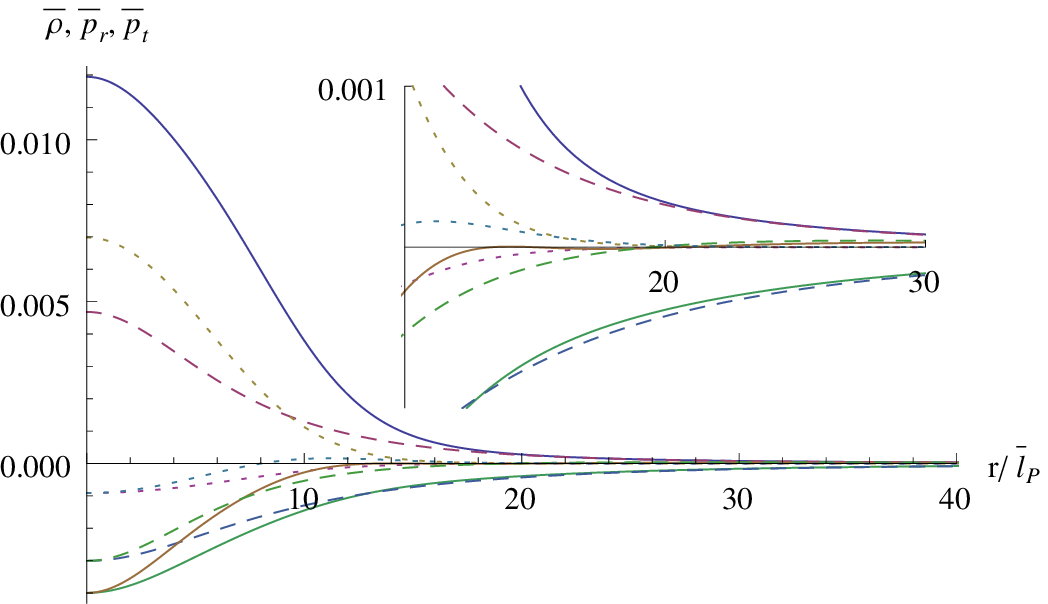}\quad
\includegraphics[scale=0.75]{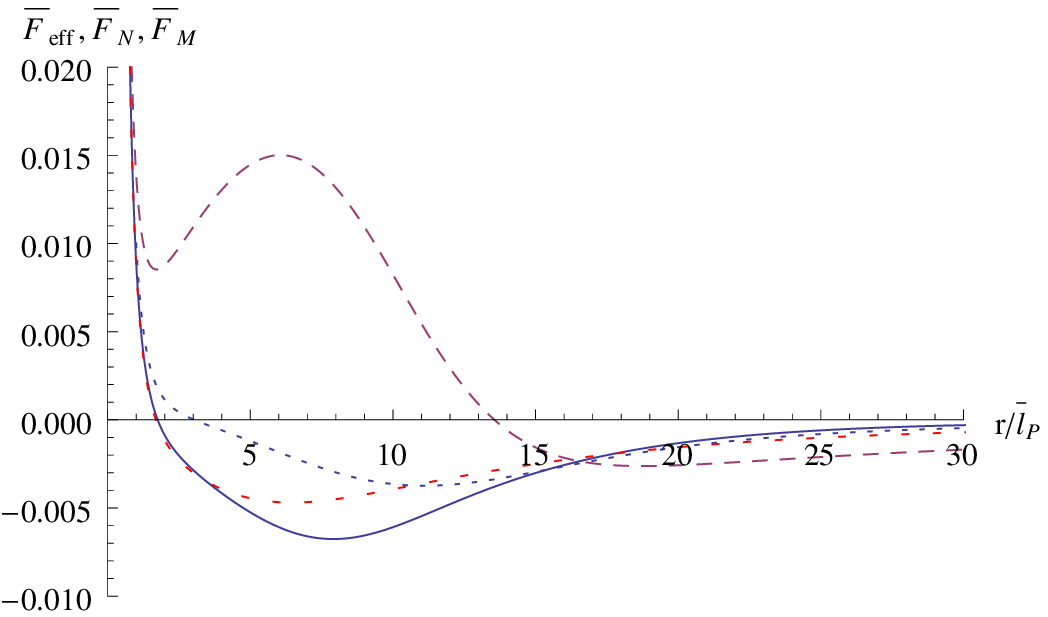}
\caption{Left plot: the energy density and pressures for the boson star alone (dotted curves), the global monopole alone (dashed curves) and the combined system of the boson star and the global monopole (solid curves). Right plot: the effective force~(\ref{Feff}) (solid curve), the Newtonian force~(\ref{NewtonF}) (dotted curve) and the Newtonian force produced by the core mass function~(\ref{NewtonFM}) (dashed  curve) for the combined system of the boson star and the global monopole; and the effective force of the boson star alone (sparse dashed red curve). The parameters are: $\tilde{\sigma}_0=0.05$, $\lambda_{\rm BS}=0$, $\xi_{\rm BS}=-4$, $\lambda_{\rm GM}=0.1$, $\Delta=0.08$, $\xi_{\rm GM}=2$.}
\label{MuTdBSGMa}
\end{figure}

 From the left plot of Fig.~\ref{MuTdBSGMa} we can trace the change in the behaviour of the energy density and pressures. In all three cases the energy density is positive while the pressures are negative functions of the radial coordinate. Observe that the size of the combined system is approximately the same as the size of the global monopole (the boson star is a bit smaller).
 Thus, even though the central energy density of the boson star dominates the energy density of the global monopole, while the size of the boson star is smaller, the combined system is approximately of the same size as the global monopole (see inset in the left plot of Fig.~\ref{MuTdBSGMa}).
In the right plot of Fig.~\ref{MuTdBSGMa} we plot the effective force~(\ref{Feff}) (solid curve), the Newtonian force~(\ref{NewtonF}) (dotted curve) and the Newtonian force produced by the core mass function~(\ref{NewtonFM}) (dashed curve)  for the combined system. We also show the effective force of the boson star alone (sparse dashed red curve).
  Even though, the global monopole produces a repulsive effective force (while the boson star produces an attractive effective force)  the effective force produced by the combined system is more attractive than in the case of the boson star alone. Note that the radius of the stable bound orbit (where $F_{\rm eff}=0$) is almost the same for the boson star and the combined object, while the global monopole alone has no such orbits. Although there is a qualitative agreement between $\bar F_{\rm eff}$ and $\bar F_{N}$ in that they both exhibit stable bound orbits, quantitatively they differ. Since $\bar F_N$ and $\bar F_M$ significantly differ from the $\bar F_{\rm eff}$ of the combined system, the nonlinear effects and the effects of the gravitational slip are significant.

\subsection{Strong coupling regime}\label{Strong}
In this subsection we examine the combined system of the boson star and the global monopole when the repulsive monopole effects are  strong. As we have seen in the previous section, this is the case for large positive $\xi_{\rm GM}$. Here we show the examples with $\xi_{\rm GM}=5$ and  $\xi_{\rm GM}=8$, while the other parameters are the same as in the previous subsection. As we shall see in what follows, the effects of strong gravitational fields increase dramatically with increasing $\xi_{\rm GM}$, reaching compactness close to unity in the latter case. Increasing $\xi_{\rm GM}$ even further leads to numerically unstable solutions which we interpret as a signature of event horizon formation and therefore black hole forms.
\begin{figure}
\centering
\includegraphics[scale=0.75]{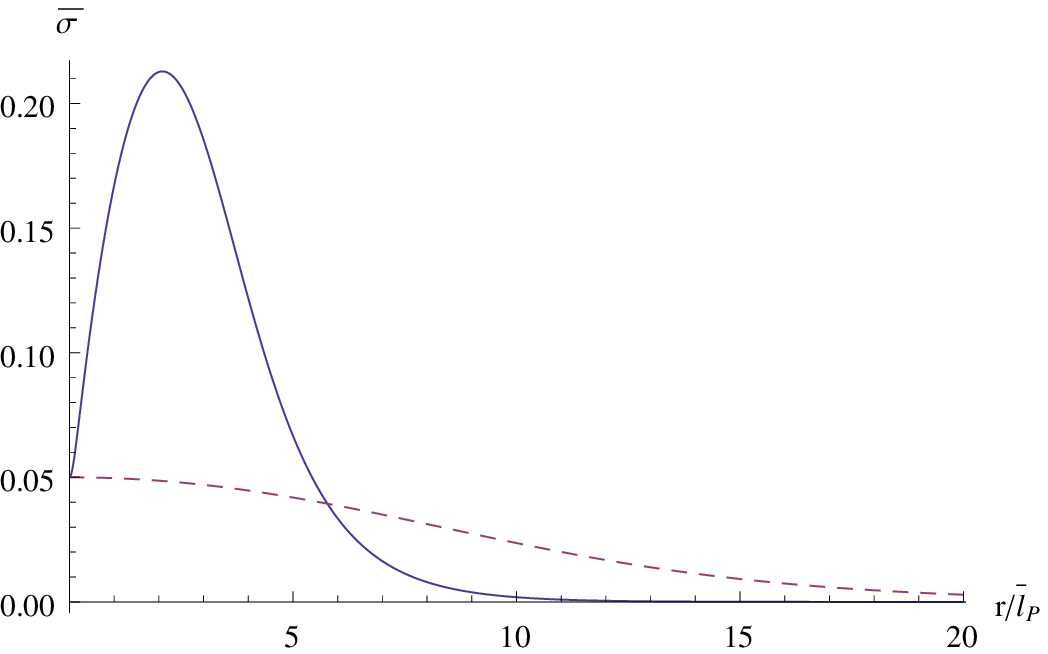}\quad
\includegraphics[scale=0.75]{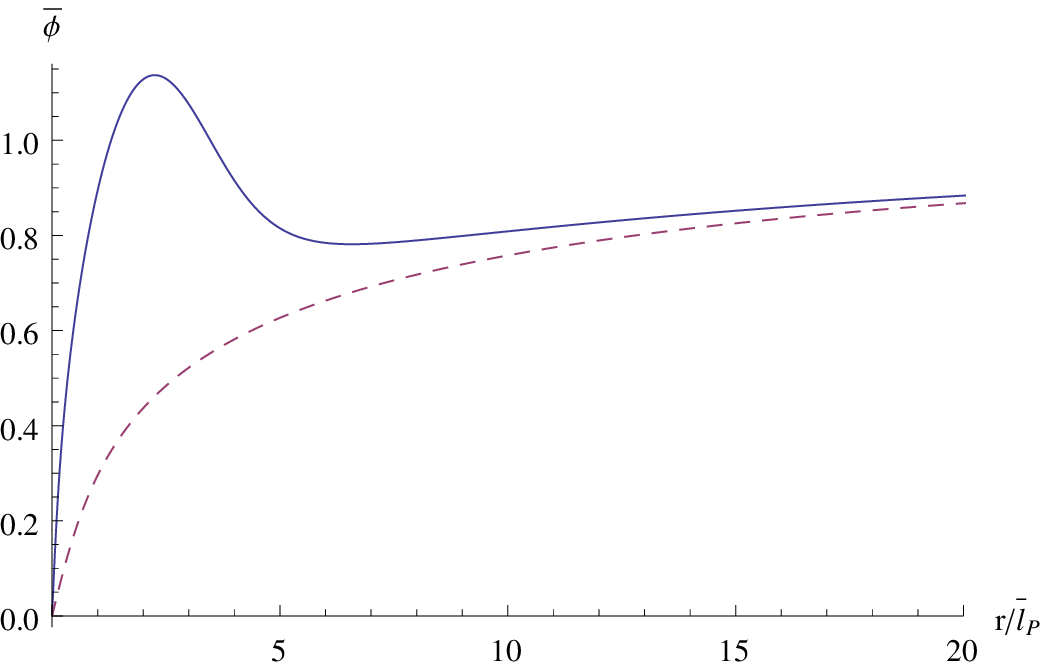}
\caption{Left plot: the boson star field alone (dashed curve) and the boson star field in the presence of the global monopole (solid curve). Right plot: the global monopole field alone (dashed curve) and the global monopole field in the presence of the boson star (solid). The parameters are: $\tilde{\sigma}_0=0.05$, $\lambda_{\rm BS}=0$, $\xi_{\rm BS}=-4$, $\lambda_{\rm GM}=0.1$, $\Delta=0.08$, $\xi_{\rm GM}=5$.}
\label{poljeBSGMd}
\end{figure}

In Fig.~\ref{poljeBSGMd} we already see that now both, the boson star field (left plot) and the monopole field (right plot) are strongly influenced by each other: the repulsive monopole  and the attractive boson star in the combined system strongly  influence individual configurations.

\begin{figure}
\centering
\includegraphics[scale=0.75]{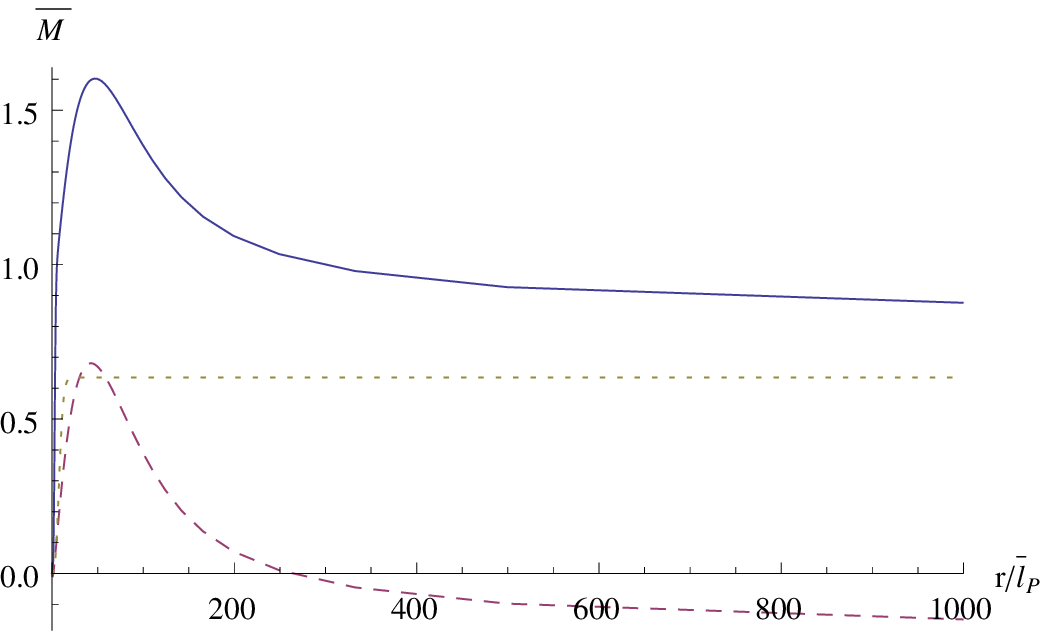}\quad
\includegraphics[scale=0.75]{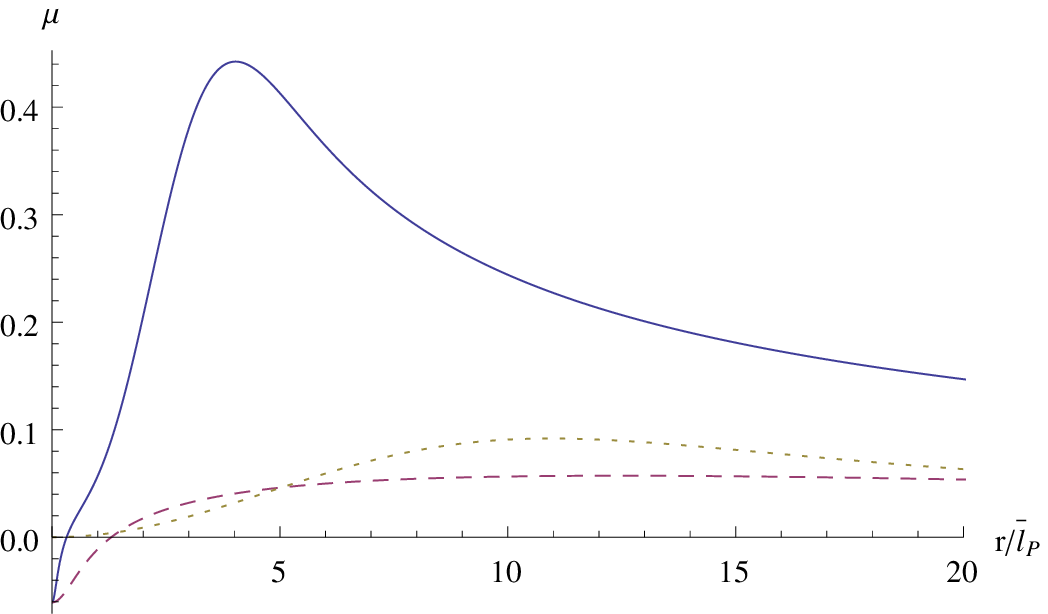}
\caption{Left plot: the mass function for the boson star alone (dotted curve), the core mass function for the global monopole alone (dashed curve) and the core mass function for the combined system of the boson star and the global monopole (solid curve). Right plot: the compactness for the boson star alone (dotted curve), the global monopole alone (dashed curve) and the combined system (solid curve). The parameters are: $\tilde{\sigma}_0=0.05$, $\lambda_{\rm BS}=0$, $\xi_{\rm BS}=-4$, $\lambda_{\rm GM}=0.1$, $\Delta=0.08$, $\xi_{\rm GM}=5$.}
\label{MaseMuBSGMd}
\end{figure}

However, the core mass function as shown in the left plot of Fig.~\ref{MaseMuBSGMd} behaves similarly as in the $\xi_{\rm GM}=2$ case while the compactness function is enlarged significantly when compared with the sum of the two, as can be seen in the right plot of Fig.~\ref{MaseMuBSGMd}.
Moreover, the maximum compactness in this case is slightly above $0.4$ thus a bit larger then the maximum compactness that can be reached in the case of (non)minimally coupled boson stars (which is about $0.32$).

\begin{figure}
\centering
\includegraphics[scale=0.75]{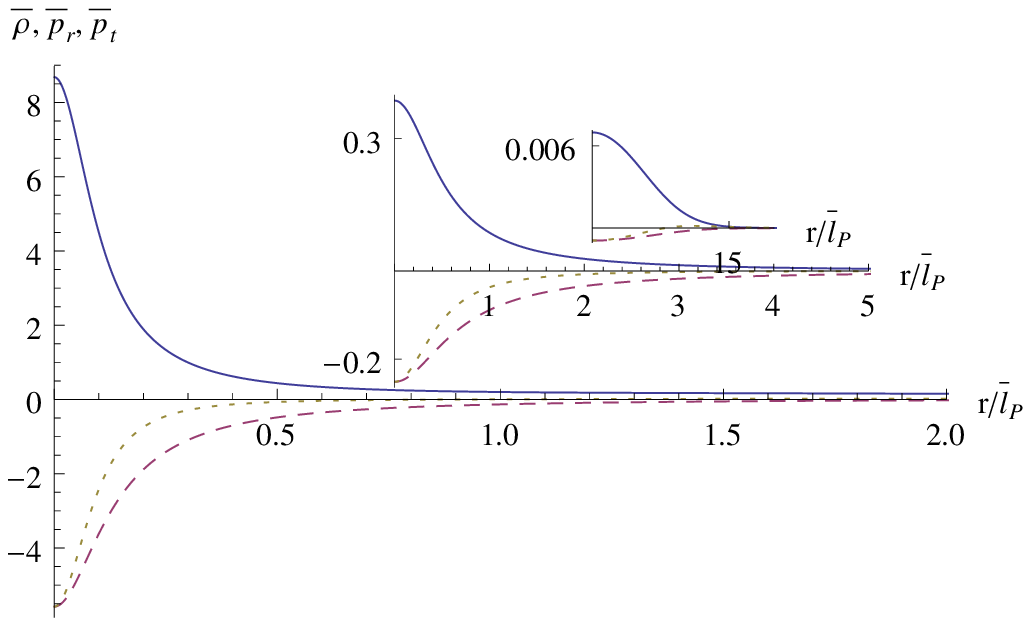}\quad
\includegraphics[scale=0.75]{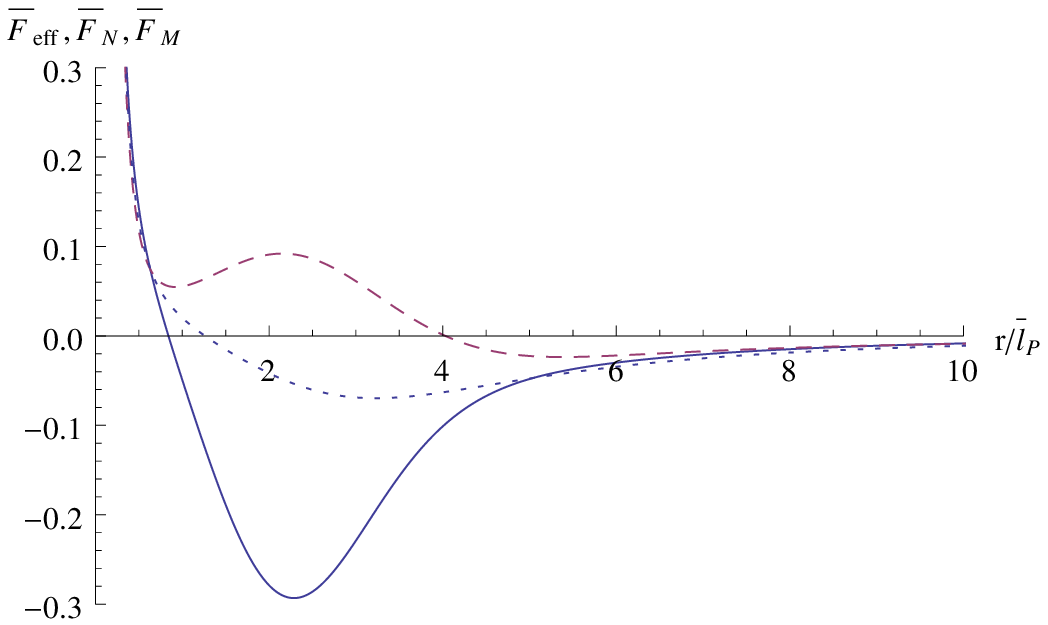}
\caption{The energy density and pressures for the boson star alone (dotted curves), the global monopole alone (dashed curves) and the combined system of the boson star and the global monopole (solid curves).  The parameters are: $\tilde{\sigma}_0=0.05$, $\lambda_{\rm BS}=0$, $\xi_{\rm BS}=-4$, $\lambda_{\rm GM}=0.1$, $\Delta=0.08$, $\xi_{\rm GM}=5$.}
\label{tdBSGMd}
\end{figure}

The whole system has shrunk as it is obvious from the left plot of Fig.~\ref{tdBSGMd} - the combined object is much smaller then its constituents.
The forces produced by the combined system are also quite strong as shown in the right plot of Fig.~\ref{tdBSGMd}. Solid curve shows  $\bar F_{\rm eff}$, dotted $\bar F_N$  and dashed $\bar F_M$. When compared with the mild coupling regime, the nonlinear gravitational effects and the gravitational slip are similar, but amplified.

\begin{figure}
\centering
\includegraphics[scale=0.8]{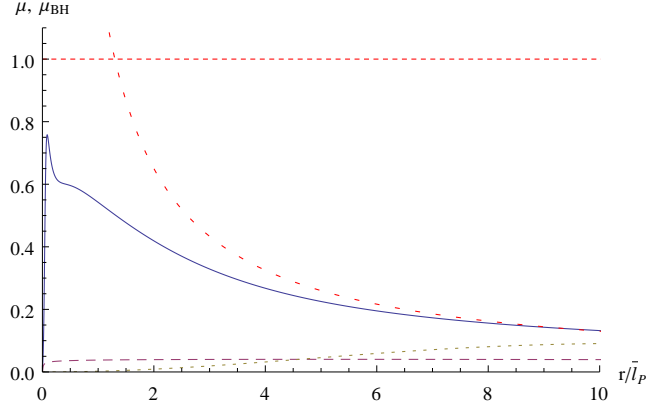}
\caption{The compactness for the combined system  of the boson star and the global monopole (solid curve), the boson star alone (dotted curve) and the global monopole alone (dashed curve). The compactness for a Schwarzschild black hole with the mass equal to the core mass of the combined system (sparse dashed red curve). The parameters are: $\tilde{\sigma}_0=0.05$, $\lambda_{\rm BS}=0$, $\xi_{\rm BS}=-4$, $\lambda_{\rm GM}=0.1$, $\Delta=0.11$, $\xi_{\rm GM}=8$.}
\label{muBSGMg}
\end{figure}

As  $\xi_{\rm GM}$ further increases the object shrinks further and the maximum compactness increases, approaching values comparable to unity which signals formation of a black hole.
To show this, in Fig.~\ref{muBSGMg} we plot the compactness for $\xi_{\rm GM}=8$, for which the maximum value is slightly above $0.75$ (solid curve).
This high value for the combined object is reached although individual maximum compactnesses are quite small: for the monopole alone
$\mu_{\rm max}\simeq 0.05$ while for the boson star alone $\mu_{\rm max} \simeq 0.1$. In order to find out whether this highly compact object can be a good black hole mimicker, we also show the compactness of a Schwarzschild black hole with a mass that corresponds to the asymptotic mass of the combined system (sparse dashed red curve). A comparison of the two curves shows that, up to a radius about a few times the Schwarzschild radius, the compactness of the black hole can be well approximated by that of the combined system. This suggests that any physical process that occurs at distances up to a few times the event horizon can be well approximated by this black hole mimicker.

\begin{figure}
\centering
\includegraphics[scale=0.8]{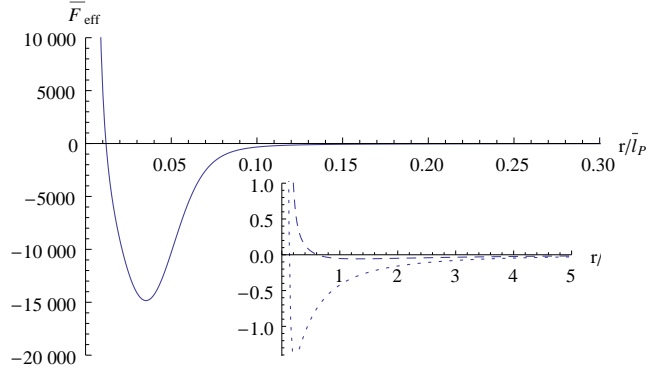}
\caption{The effective force~(\ref{Feff}) of the combined system of the boson star and the global monopole. Inset: the Newtonian force (dotted curve) and the Newtonian force produced by the core mass function (dashed curve). The parameters are: $\tilde{\sigma}_0=0.05$, $\lambda_{\rm BS}=0$, $\xi_{\rm BS}=-4$, $\lambda_{\rm GM}=0.1$, $\Delta=0.11$, $\xi_{\rm GM}=8$. Also the angular momentum and the energy (of the particle) per unit mass are $\bar L=0.1$ and $\bar E=1$.}
\label{sileBSGMg}
\end{figure}

Of course there is a prize to pay, and there are extremely large effective forces that are developed in the vicinity of the radius where the compactness maximizes, as can be seen in Fig.~\ref{sileBSGMg}. Indeed, the effective force reaches an extremely large value above 10000, whereby
there are no large numbers present in any of the couplings. This can be explained by the large compression exerted on the boson star by the monopole gravitational field. In this process a crucial role is played by the gravitational backreaction as well as by nonlinear effects. This can be seen from the inset in Fig.~\ref{sileBSGMg}, where we show the forces $\bar F_N$ (dotted curve) and  $\bar F_M$ (dashed curve) which are of the order of unity in the relevant region, thence tremendously different from $\bar F_{\rm eff}$. Increasing $\xi_{\rm GM}$ further above $8$ leads to a further dramatic
increase in the compression of the boson star and the effective force, signaling gravitational instability and formation of a black hole.
While we have here managed to form a fine black hole mimicker, the prize was a tuning in the parameters. Namely, for each choice of the coupling
there is a critical value of the nonminimal coupling $\xi_{\rm GM}$, above which a black hole forms, and below which a highly compact object forms
with properties close to a black hole.

\section{Conclusions and discussion}\label{Concl}

 In this paper we have considered the compact object composed of a nonminimally coupled boson star and a nonminimally coupled monopole.
 Three distinct regimes have been identified: weak, mild and strong coupling regimes. The main parameter that determines the regime is
the nonminimal coupling of the global monopole.

 In the weak coupling regime (when, \emph{e.g.}, $\xi_{\rm GM}= -1$) nonlinear gravitational effects and the gravitational
backreaction are weak, and the resulting compact object can be obtained by summing the energy densities and pressures of the two components.

In the mild coupling regime (when, \emph{e.g.}, $\xi_{\rm GM}= 2$) we have seen that the nonlinear effects and the effects of gravitational slip are present and the resulting object behaves as a boson star with a larger compression and thus with a larger compactness.

In the strong coupling regime (when,  \emph{e.g.}, $\xi_{\rm GM}\gtrsim 5$) a large compression of the composite object takes place such that when $\xi_{\rm  GM} \sim 8$ one can get a highly compact object with the maximum compactness of the order unity. This object represents a good black hole mimicker in that, up to distances close to the black hole event horizon, the compactness profile of the mimicker follows closely that of the black hole. For even larger values of $\xi_{\rm GM}$  we do not get stable configurations. We interpret it as a signal for black hole formation.\\
It would be of interest to investigate the stability of these highly compact and dense objects, and $M(R)$ stability analysis seems a natural method.  We wish to point out that the stability analysis needs to be done with a due care, since boson stars get largely compressed in the
presence of a global monopole, and increasing the boson star mass may lead to a more compact but still stable object. A naive application of $M(R)$ method would suggest instability, while in reality the object may be stable. These thoughts suggest not only the need for a proper stability analysis, but also that it may require a nontrivial modification of standard methods.
 The result of the last subsection in this paper indicates that it is the global monopole that stabilizes the monopole-boson star composite
 system against collapse. Recall that global monopoles are classical field configurations stabilized by topology, while
  boson stars are stabilized by scalar current density, or equivalently by scalar field charge. With this in mind we make the following conjecture:
\begin{itemize}
\item[]   \emph{Compact star objects stabilized by a global charge tend to be more stable than those stabilized by a (local) charge density, and hence are
 better black hole mimickers.}
 \end{itemize}
 In addition to performing a detailed stability analysis, it would be useful to perform a detailed analysis to what extent are
 the objects composed of a global monopole and a boson star good black hole mimickers. Here we have only compared in some detail
 the compactness profiles of the mimickers with those of a true black hole. But of course, there are further comparisons one should investigate, and these include: a detailed comparison of bound (stable and unstable) orbits; the creation and emission of gravitational waves in binary star systems
 (in which one or two companions is a black hole mimicker); vibrational modes (\emph{i.e.} modes that govern deviations from spherical symmetry) and their decay rates, \emph{etc}. Once such studies are complete we will have a much better idea on to what extent the dense compact objects considered in  this paper are good black hole mimickers.

\section*{Acknowledgements}
A.M. is grateful for the hospitality of the Institute for Theoretical Physics at Utrecht University where part of this work was carried out. This work is partially supported by the Croatian Ministry of Science under the project number 036-0982930-3144.

\section*{Appendix A: Newtonian limit of the Einstein equations}\label{C}
In the Newtonian limit, \emph{i.e.} for $c\rightarrow\infty$, the metric line element is given with (see \emph{e.g.} Ref.~\cite{LL})
\beq
ds^2=-(1+2\phi_N)dt^2 + (1-2\psi_N)d\vec{r}^{\,2},
\label{metricNewton}
\eeq
where $\phi_N=\psi_N$ is the Newtonian potential. From the Einstein equation
\beq
G_{\mu}^{\nu}=R_{\mu}^\nu-\frac 12 g_\mu^\nu R=8\pi G_N T_\mu^\nu
\eeq
we can express Ricci scalar $R$ in terms of the energy-momentum scalar $T$ via the trace equation
\beq
R=-8\pi G_N T
\eeq
leading to another form of the Einstein equations
\beq
R_\mu^\nu=8\pi G_N\left(T_\mu^\nu-\frac 12 g_\mu^\nu T\right).
\eeq
For the given metric~(\ref{metricNewton}) the only non-vanishing component of the Riemann tensor is $R_0^0$, thus we have only one Einstein equation in the Newtonian limit
\beq
R_0^0=4\pi G_N(T_0^0-T_i^i).
\eeq
Inserting~(\ref{metricNewton}) in the formula for the Riemann tensor we arrive at
\beq
R_{00}=-\frac{\partial\Gamma_{00}^\alpha}{\partial x^\alpha}=\Delta\phi_N.
\eeq
If we now recall that the energy-momentum tensor for an anisotropic fluid is
\beq
T_\mu^\nu=\mbox{diag}(-\rho,p_r,p_t,p_t)
\eeq
we arrive at the Poisson equation for the Newtonian potential
\beq
\Delta\phi_N=4\pi G_N(\rho+\sum p_i).
\eeq
If we now use the following identity ($\phi_N=\phi_N(r)$):
\beq
\Delta\phi_N=\frac{1}{r^2}\left(r^2 \phi_N^\prime\right)^\prime,
\eeq
where the primes denote derivatives with respect to $r$, it follows that the Newtonian force is
\beq
F_N=-\phi_N^\prime=-\frac{1}{r^2}\int_0^r 4\pi G_N (\rho+\sum p_i)\tilde r^2 d\tilde r.
\eeq
Comparing this expression with the standard Newton law, we can read off  the \emph{active gravitational mass}
\beq
M(r)=4\pi \int_0^r (\rho+\sum p_i)\tilde r^2 d\tilde r.
\label{agm}
\eeq
The Newtonian force felt by a test particle with an angular momentum (per unit mass) $L$ is:
\beq
F_N=-G_N\frac{M(r)}{r^2}+\frac{L^2}{r^3},
\label{NewtonF}
\eeq
where $M(r)$ is the active gravitational mass given with the~Eq. (\ref{agm}).

%
%

\section*{Appendix B: Bound orbits}\label{D}
The geodesic equation
\beq \frac{d^2
x^\mu}{d\tau^2}+\Gamma^\mu_{\alpha\beta}\frac{dx^\alpha}{d\tau}\frac{dx^\beta}{d\tau}=0.
\label{geodesic eqBO} \eeq
in the metric
\beq ds^2=-e^\nu dt^2+e^\lambda dr^2+r^2 (d\theta^2+\sin^2\theta d\varphi^2).
\label{metricBO}\eeq
can be written in terms of its components ($x^\mu=t$ and $x^\mu=\varphi$)
\bea
&&\frac{d^2t}{d\tau^2}+\nu^\prime\frac{dt}{d\tau}\frac{dr}{d\tau}=0,\\
&&\frac{d^2\varphi}{d\tau^2}+\frac{2}{r}\frac{d\varphi}{d\tau}\frac{d
r}{d\tau}+2\frac{\cos\theta}{\sin\theta}\frac{d\theta}{d\tau}\frac{d\varphi}{d\tau}=0,
\eea
from which we extract the two Killing vectors:
\bea K_\mu &=&(-e^\nu,0,0,0),\\
F_\mu&=&(0,0,0,r^2\sin^2\theta),
 \eea
which lead to the conserved quantities (see
\emph{e.g.}~\cite{Carroll}):
\bea
E&=&-K_\mu
\frac{dx^\mu}{d\tau}=e^\nu\frac{dt}{d\tau},\quad\mbox{\emph{i.e.}}\quad
\frac{d}{d\tau}E=0,\\
L&=&F_\mu
\frac{dx^\mu}{d\tau}=r^2\sin^2\theta\frac{d\varphi}{d\tau},\quad\mbox{\emph{i.e.}}\quad
\frac{d}{d\tau}L=0.
\eea
Here $E$ is the conserved energy (per unit mass) and $L$ is the
conserved angular momentum (per unit mass). Since the direction of the angular
momentum is conserved, without loss of generality we can set $\theta=\pi/2$. From the
velocity normalization condition $u_\mu u^\mu=-1$ and making use
of the conserved quantities $E$ and $L$ we obtain an expression
for the radial velocity squared:
\beq
\left(\frac{dr}{d\tau}\right)^2=e^{-\nu-\lambda}E^2-e^{-\lambda}\left(\frac{L^2}{r^2}+1\right).
\eeq
If we rewrite this equation in a slightly different form
\beq
\frac 1 2
\left(\frac{dr}{d\lambda}\right)^2+V_{\rm eff}(r)=\mathcal{\varepsilon},
\eeq
where $\mathcal{\varepsilon}=E^2/2$, the effective potential
$V_{\rm eff}(r)$ can be read off:
\beq
V_{\rm eff}(r)=\frac{e^{-\lambda}}{2}\left(\frac{L^2}{r^2}+1\right)+\frac{E^2}{2}\left(1-e^{-\nu-\lambda}\right)
\label{eff potentialBO}
\eeq
The effective force is then
\beq
F_{\rm eff}=-\nabla V_{\rm eff}.
\label{Feff}
\eeq
If the effective potential exhibits a local minimum (node in the effective force) in the radial coordinate then stable bound orbits are possible. If on the other hand, the effective potential has a local maximum (also a node in the effective force) then bound orbits are unstable.

%

\end{document}